\documentclass[a4paper]{article}
\usepackage{fullpage}
\usepackage[latin1]{inputenc}
\usepackage{amsmath}
\usepackage{graphicx}
\usepackage{hyperref}
\usepackage{color}
\usepackage[english]{babel}

\title{Stochastic perturbation of integrable systems: a window to weakly
chaotic systems.}
\author{Khanh-Dang Nguyen Thu Lam and Jorge Kurchan}
\date{\today}

\newcommand{\red}[1]{{\color{red} #1}}

\begin{document}
\maketitle

\begin{abstract}
Integrable non-linear Hamiltonian systems perturbed by additive  noise
develop a Lyapunov instability, and are hence chaotic, for any
amplitude of the perturbation. This phenomenon is related, but distinct,
from  Taylor's diffusion in hydrodynamics.  We develop expressions for
the Lyapunov exponents  for the cases of white and colored noise.  The
situation described here being `multi-resonance' -- by nature well
beyond the Kolmogorov-Arnold-Moser  regime, it offers an analytic
glimpse on the regime in which many near-integrable systems, such as
some planetary systems,  find themselves  in practice.  We show with the
aid of a simple example, how one may model in some cases weakly chaotic
deterministic systems by a stochastically perturbed one, with good
qualitative results.
\end{abstract}

\section{Introduction}

\vspace{.5cm}

{\bf The problem}

\vspace{.5cm}

Lyapunov exponents measure the average rate of expansion of volumes
advected by the trajectory of a dynamical system.  When a dynamical
system is chaotic, some of its Lyapunov exponents are positive, a small
difference in the initial conditions is amplified exponentially with
time. An integrable system with $N$ degrees of freedom, having $N$
constants of motion, has all its  Lyapunov equal to zero. The motion is
restricted to an $N$-dimensional torus in $2N$-dimensional phase-space.

Consider one such integrable Hamiltonian dynamics, but now perturbed by
a {\em weak}  additive noise:
\begin{equation}
\begin{aligned}
\dot q_i &= \frac{\partial H}{\partial p_i} \\
\dot p_i &= -\frac{\partial H}{\partial q_i}  + \varepsilon^{1/2}\xi_i(t)
\end{aligned}
\label{hamilton1}
\end{equation}
In this paper we shall mostly consider the case in which the  $\xi(t)$
are independant gaussian white noises: 
\begin{equation}
\langle \xi_i(t) \rangle = 0
\qquad
\text{and}
\qquad
\langle \xi_i(t) \xi_i(t') \rangle = 2 \delta_{ij} \delta(t-t') .
\end{equation}
Such a system diffuses slowly from one torus to another, but we shall
consider times short enough that this diffusion is small.

It may come as a surprise that {\em for every $\varepsilon>0$} the
system (\ref{hamilton1}) generically develops a Lyapunov instability:
two trajectories starting at nearby points and {\em subjected to the
same noise} ${\bf \xi}$ diverge exponentially  (mostly, as we shall see,
on the surface of the torus): the system acquires $N$ positive Lyapunov
exponents.  Because the underlying Hamiltonian system is, by assumption,
integrable, the exponents vanish in the limit of zero noise amplitude
-- as $\varepsilon^{1/3}$, as we show below \cite{celia}. In what follows we
shall derive expressions for this, and more general situations.


\vspace{.5cm}

{\bf Motivation}

\vspace{.5cm}

Before launching into rather long calculations, let us discuss our
motivation.  Systems that are integrable and subjected to a small
{non-integrable} perturbation are quite common in physics: the
example of planetary systems, where the perturbation is the interaction
between different planets, immediately comes to mind. Another family of
problems of this kind arises when one considers a system with many
interacting degrees of freedom $N$, such that in some initial condition
the average interaction is integrable in the $N \rightarrow \infty$
limit.  This is the case of stars belonging to a (to a first
approximation) homogeneous, spherical stellar cloud: each star perceives
the rest as a spherical integrable potential, although the system is
most definitely not integrable when one takes into account the
inhomogeneities of mass distribution.
{Finally, one should remark that even numerical roundoff errors
themselves may induce a Lyapunov instability in a system that has none,
at least when the Lyapunov exponents are calculated on the basis of the
tangent dynamics associated with a single trajectory.}

Small perturbations of integrable systems  evoke the
Kolmogorov-Arnold-Moser (KAM) theorem, which states that under certain
conditions, once perturbation is turned on, regularity is not totally lost, and
there remain some regions where trajectories belong to tori and have zero
Lyapunov exponents.  Remarkable as it is, the KAM result is very often
irrelevant as soon as one considers systems with a few degrees of freedom.
Indeed, planetary systems such as the solar system are known to be
chaotic~\cite{laskar1989,wisdom1987}. Even more dramatically, nonlinear chains
of springs (the Fermi-Pasta-Ulam problem) are expected~\cite{Livi} to be
regular only at temperatures exponentially small in the chain length.  The
reason for the fragility of the KAM regime is easy to understand: a regular
region in phase space requires that every degree of freedom be regular, just
any subsystem becoming chaotic would spoil the regularity of  the rest -- {\em
an expectation made more plausible by the result  in this paper  that a (random)  perturbation
of arbitrarily small amplitude   renders a system chaotic}.  Hence,
one may estimate  that the size of regular islands is a {multiplicative
process} that scales exponentially with the dimension.  A regime of stronger
chaoticity have been discussed  by Nekhoroshev~\cite{Morbi}, where Lyapunov
exponents are non-zero, but exponentially small in the (perturbation)$^{-1}$.
The situation we discuss here is even beyond that, and it corresponds to a
situations where there are many resonances of all frequencies~\cite{Morbi}.

As mentioned above, the stochastic perturbation which is our main
concern here,  drives the system out of regularity even for arbitrarily
small amplitudes. To understand that this does not contradict the KAM
theorem, we argue as follows: the stochastic equation (\ref{hamilton1}),
a Langevin process with {infinite temperature}, may be derived by
considering the system coupled with a bath composed of an infinite
number of oscillators, with a continuum spread of
frequencies~\cite{Zwanzig}. We are hence in a situation as described
above: we may think of (\ref{hamilton1}) as a system with  infinitely
many degrees of freedom, those of the original system plus those of the
bath. 

The purpose of this paper is then to understand in better detail this
regime, as a basis for treating systems in which the perturbation is not
stochastic, but is due to the effect of the rest of the system with a
particular degree of freedom.

\vspace{.5cm}

{\bf The Lyapunov and the Taylor  regimes}

\vspace{.5cm}

Before concluding this introduction, let us write Equations
(\ref{hamilton1}) in a more flexible and general way.  Considering a
bath of oscillators coupled in a generic way to a Hamiltonian  system,
one  may write   the most general Markovian Langevin equation (here
restricted to the infinite temperature limit) in a canonically invariant
way~\cite{cepas1998}.  Denoting  $G_k$ phase-space functions  which
specify the coupling between system an bath, and the phase-space
variables  ${\bf x} = (q_1,\dots,p_1,\dots)$, one has:
\begin{equation}
\begin{aligned}
\red{(S)}\qquad
\dot x_i &= \{x_i,H\} + \varepsilon^{1/2} \sum_k \{x_i,G_k\}\xi_k(t) \\
\red{(I)}\qquad
\dot x_i &= \{x_i,H\} +  \varepsilon^{1/2} \sum_k \{x_i,G_k\}\xi_k(t)
    + \varepsilon \sum_k \{G_k,\{G_k,x_i\}\}
\end{aligned}
    \label{langevin1}
\end{equation}
Here $\{A,B\}$ are the Poisson brackets.
The first is the equation in the Stratonovitch, and the second in the
Ito convention.
One can now check that the usual Langevin equations (\ref{hamilton1})
are obtained for $G_k= -q_k$.  Equation (\ref{langevin1})  leads to the evolution for the
phase-space probability distribution $P({\bf x})$:
\begin{equation}
\frac{\partial P}{\partial t} + \{P,H\} = 
\varepsilon\sum_k   \bigl[ \{G_k, \{G_k,H\}P\}
+ \{G_k, \{G_k,P\}\} \bigr]
\end{equation}
We have made explicit the amplitude $\varepsilon$ of the noise, which we
shall assume throughout  to be small.

The advantage of this canonically invariant representation is that we
may  take advantage of the integrable nature of the Hamiltonian
dynamics: we may now write everything in terms of the angle $\theta_i$
and action $I_i$ variables. The Hamiltonian is then a function of the
$I_i$, and the equations (\ref{langevin1}) read, for example in the
Stratonovitch convention: 
\begin{align*}
\qquad
    \dot I_i &=  \varepsilon^{1/2} \sum_k \{I_i,G_k\}\xi_k(t) \\
\qquad
    \dot \theta_i &= \omega_i +  \varepsilon^{1/2}\sum_k \{\theta_i,G_k\}\xi_k(t)
   \label{langevin2}
\end{align*}
Here, $\omega_i = \frac{\partial H}{\partial I_i}$ is the angular frequency of $\theta_i$.
The $G_k$ have  to be expressed in terms of the action and angle
variables, 
$\{\theta_i,G_k\} = \frac{\partial G_k}{\partial I_i}$
and
$\{I_i,G_k\} = -\frac{\partial G_k}{\partial \theta_i}$.

In the absence of noise, the system remains confined to a torus labeled
by the value of the constants of motion $I_i$, and spanned by the
$\theta_i$. The effect of the noise is to add some diffusion, within and
away from the torus.  Because the amplitude of the noise is by
assumption small (of order $\varepsilon^{1/2}$) and random, the
typical time for this diffusion is $t_\text{diff}\sim \varepsilon^{-1}$.
Consider now two trajectories starting in nearby points on the same
torus, under the effect of the same noise: apart from their common
diffusion, there is an exponential separation of trajectories that, as
we shall see, has a characteristic time $\tau \sim
\varepsilon^{-1/3}$.  
Once two trajectories have diverged substantially (of $O(1)$), the  fact
that their noise is the same becomes irrelevant, and each follows its
own diffusion.  In the small $\varepsilon$ limit, there is a large range
of timescales  $\varepsilon^{-1/3} \ll \tau  \ll \varepsilon^{-1}$
where the  diffusive drift away from a torus is still very small, but
the Lyapunov instability is well defined. We shall in what follows
concentrate on such times.

The interplay of noise and regular dynamics has a history in the
hydrodynamics of laminar flows: the enhancement  in diffusion due to the
interplay with regular advection goes under the name of Taylor
diffusion.  The effect we study here is related but distinct, and it
corresponds to a different regime.   In order to better understand this,
consider the following example:
\begin{equation}	\label{eq:poiseuille}
H=  I - \frac{1}{6} I^3   \qquad ; \qquad \omega(I)= 1 - \frac12 I^2,
\end{equation}
and let us choose:
\begin{equation}	\label{eq:poiseuilleG}
G = \sqrt{2}\cos\theta.
\end{equation}
The equations of motion read:
\begin{equation}	\label{eq:poiseuille:eqmvt}
(I)\qquad
\begin{aligned}
\dot\theta &= \omega(I) \\
\dot I &= - (2 \varepsilon)^{1/2}  \sin \theta  \; \xi(t)
\end{aligned}
\end{equation}
As we shall see below,  because $\sin \theta$ is multiplying a white
noise of small amplitude in (\ref{eq:poiseuille:eqmvt}), it may
without loss of generality be replaced by its root-mean-square average.
If we now make the identification of $I$ as the transverse and $\theta$
the longitudinal direction, our example corresponds  then precisely to a
Poiseuille flow on a two-dimensional channel~\cite{guyon}, with
transverse  diffusion -- the textbook example of Taylor diffusion:
 \begin{equation}	\label{eq:poiseuille:eqmvt1}
(I)\qquad
\begin{aligned}
\dot x &= 1-y^2 \\
\dot y &= - {\varepsilon}  \; \xi(t)
\end{aligned}
\end{equation}

\begin{figure}
\begin{center}
\includegraphics{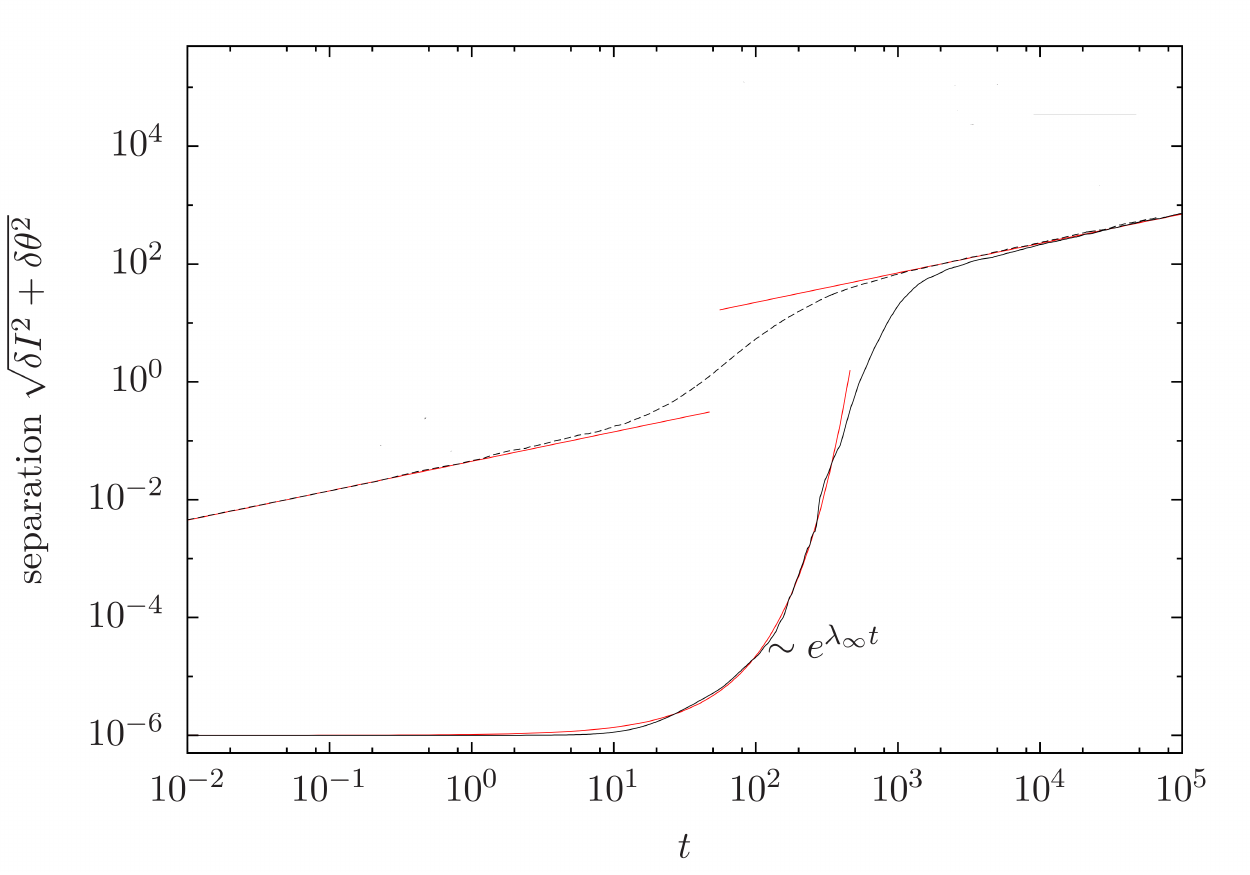}\\
\includegraphics{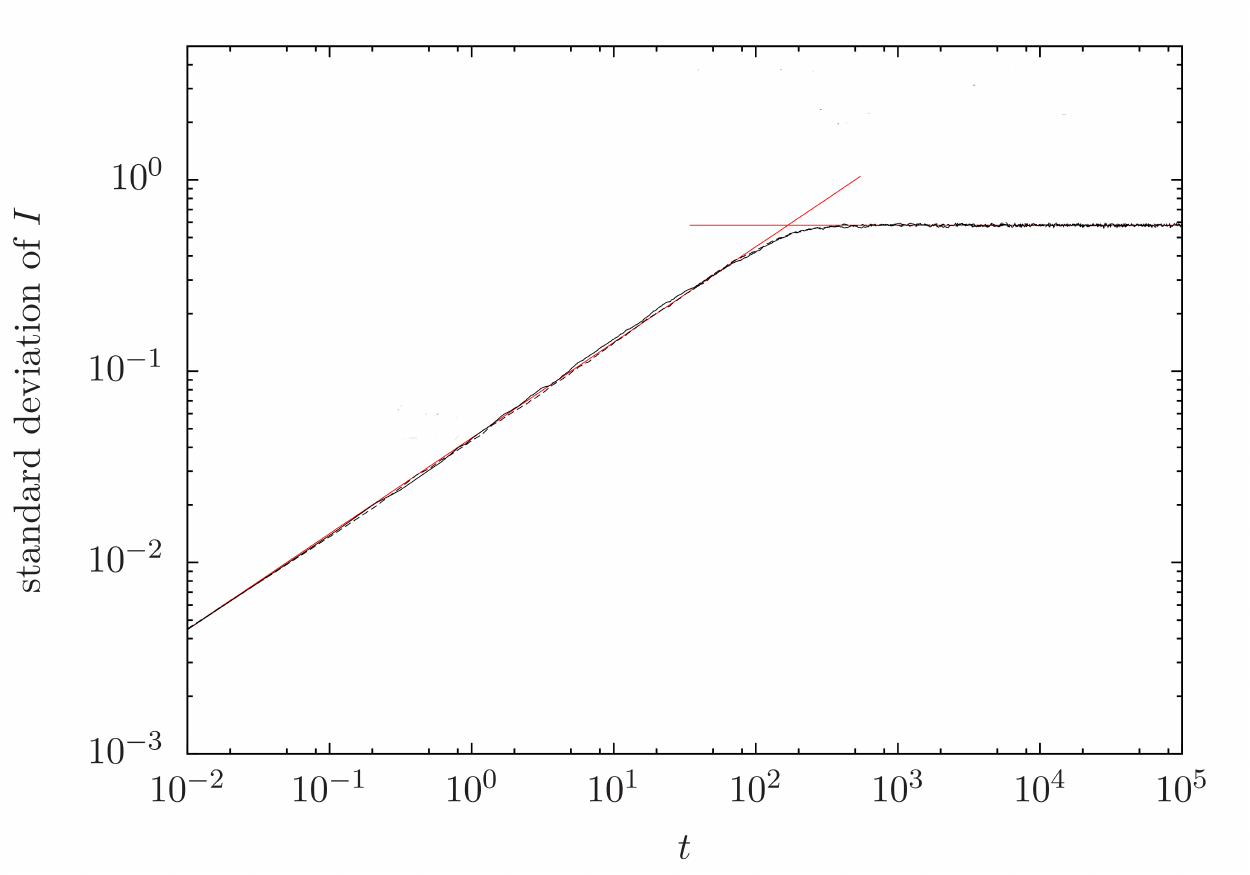}
\end{center}
\caption{
\label{fig:poiseuille}
(a)~Evolution of the distance between  two points at $I=0$  initially separated
by a distance $\sqrt{\delta I^2 + \delta\theta^2} = 10^{-6}$.  We let the
systems evolve with Eqs.\eqref{eq:poiseuille:eqmvt}, with diffusion
coefficient $\varepsilon=10^{-3}$. The results are averaged over $1024$
realizations.  Dotted line and full line correspond to different and same
noise for the two realizations.  For different noise realizations there is an initial diffusion with coefficient $\varepsilon$, followed by a faster Taylor diffusion with $\varepsilon_{eff}=0.0025/\varepsilon$
For the case of equal noise there is an initial  exponential separation, followed
 by the  Taylor
diffusion regime with $\varepsilon_{eff}$.
(b)~Displacement along $I$ for the same problem.
}
\end{figure}

%
The results  may be seen in figure~\ref{fig:poiseuille}.
When the two particles have independent realizations of
noise, their separation evolves in a purely diffusive manner, as
$\sqrt{2\varepsilon t}$, until the diffusion reaches the walls, when the
distribution becomes stationary.  A surprising phenomenon occurs then:
the copies perform an essentially longitudinal diffusion with an
enhanced effective coefficient $\varepsilon_\text{eff}$. This is the
Taylor-Aris dispersion~\cite{taylor1953,aris1956}.  The origin of this
enhancement is simple: particles behave like cars which advance
deterministically along a highway with lanes having different speeds,
but diffuse laterally.  As they diffuse back to their original lane,
they do so with a fluctuation in the longitudinal direction that is the
result of the stochastic excursion along faster and slower lanes.
In Figure \ref{fig:poiseuille} one may fit $\varepsilon_\text{eff} = 0.0025/\varepsilon$, which is in
agreement with the expressions~\cite{taylor1953,aris1956} for a case
initially without diffusion along the channel.

Consider now our case, when the two particles are subjected to the same
noise.
The separation is initially exponential  $e^{\lambda_\infty t}$, where
$\lambda_\infty$ is by definition the  Lyapunov exponent.  Just as in
the case of independent noise realizations, at long times the system
crosses over to a (predominately longitudinal) Taylor dispersion regime.
In this paper we shall be mostly concerned with the initial exponential
separation regime, with copies subjected to the same noise.

\section{Evolution of the tangent vectors}

We now turn to the evolution of two nearby trajectories  $x_i(t)$ and
$x'_i(t)$, and the tangent vector
$u_i(t)= \frac{\|x_i(t)-x'_i(t)\|}{\|x_i(0)-x'_i(0)\|}$.
One has to be careful about the prescription. One obtains in the
Ito convention:
\begin{align}
\qquad
\dot x_i &= \{x_i,H\} + \varepsilon^{1/2} \sum_k \{x_i,G_k\}\xi_k(t) \nonumber
\\
\qquad
\dot u_i &= 
	\sum_j \frac\partial{\partial x_j}
	\left[
	    \{x_i,H\} + \varepsilon^{1/2}\sum_k \{x_i,G_k\}\xi_k(t)
	\right]
	u_j  \label{dos}
\end{align}

Note that the evolution of the $u_i$ is `slaved' to that of the $x_i$.
For small $\varepsilon$ and  times smaller than $t_\text{diff} \sim
\varepsilon^{-1}$, we may neglect the effect of noise on the evolution
of the $x_i$, so that the original variables  move on a torus. Further
progress is made by writing Eqs \eqref{dos} in angle-action variables
$(\theta_i,I_i)$. 
We have:
\begin{equation}	\label{hami}
\begin{aligned}
\dot \theta_i &= \{\theta_i,H\} = \omega_i({\bf I})
&\qquad\qquad
\theta_i &= \omega_i t + \omega_i^0
\\
\dot I_i &= 0
&\qquad\qquad
I_i &= I_i^0
\end{aligned}
\end{equation}
Denoting the set 
$\{u_1,...,u_{2N}\}=\{(u_{I_1},\dots, u_{I_N}),(u_{\theta_1},\dots ,u_{\theta_N})\}$,
the evolution in the tangent space becomes:
\begin{equation}
\begin{aligned}
\dot u_{\theta_i} &= 
	\sum_j
	\left[
	  \varepsilon^{1/2} \sum_{k}
	  \frac{\partial{\{\theta_i,G_k\}}}{\partial \theta_j}\xi_k(t)
	\right]
	u_{\theta j}
	+
	\sum_j
	\left[
	\frac{\partial{  \{\theta_i,H\}}}{\partial I_j}+
	 \varepsilon^{1/2} \sum_{k} \frac{\partial{\{I_i,G_k\}}}{\partial I_j}\xi_k(t)
	\right]
	u_{I_j} 
\\
\dot u_{I_i} &= 
	\sum_j
	\left[
	   \varepsilon^{1/2} \sum_{k}   \frac{\partial{\{I_i,G_k\}}}{\partial \theta_j}\xi_k(t)
	\right]
	u_{\theta j}
	+
	\sum_j
	\left[
	  \varepsilon^{1/2} \sum_{k} \frac{\partial{\{I_i,G_k\}}}{\partial I_j}\xi_k(t)
	\right]
	u_{I_j} 
\end{aligned}
\label{tres}
\end{equation}
where we may assume (for the times that concern us here) that  the
phase-space variables are unperturbed by the noise and are given by
Eqs~\eqref{hami}.

In order to compare the terms in the limit of small $\varepsilon$, we
propose a rescaling of the $u_j$ and time, as follows:
\begin{equation}
t \rightarrow \varepsilon^{-\alpha} t  \qquad ; \qquad
u_{\theta_i} \rightarrow \varepsilon^{-\beta} u_{\theta_i}   \qquad ; \qquad
u_{I_i} \rightarrow u_{I_i}
\label{uuy}
\end{equation}
We shall assume and check that $\alpha>0$ and $\beta>0$.
Equations~\eqref{tres} become:
\begin{align}
\begin{split}
\dot u_{\theta_i} &=
	\sum_j
	 \left[
	 \varepsilon^{\frac12-\alpha} \sum_{k}
	  \frac{\partial{\{\theta_i,G_k\}}}{\partial \theta_j}\xi_k(\varepsilon^{-\alpha}t)
	\right]
	u_{\theta_j}
\\
& \qquad\qquad\qquad +
	\sum_j
	\left[
	  \varepsilon^{-\alpha+\beta}
	  \frac{\partial{  \{\theta_i,H\}}}{\partial I_j}
	  +
	 \varepsilon^{\frac12-\alpha+\beta} \sum_{k}
	 		\frac{\partial{\{I_i,G_k\}}}{\partial I_j}\xi_k(\varepsilon^{-\alpha}t)
	\right]
	u_{I_j}  \label{cuatro}
\end{split}
\\
\dot u_{I_i} &=
	\sum_j
	\left[
	   \varepsilon^{\frac12-\alpha-\beta} \sum_{k}
	   	\frac{\partial{\{I_i,G_k\}}}{\partial \theta_j}\xi_k(\varepsilon^{-\alpha}t)
	\right]
	u_{\theta_j}
	+
	\sum_j
	\left[
	  \varepsilon^{\frac12-\alpha} \sum_{k}
		\frac{\partial{\{I_i,G_k\}}}{\partial I_j}\xi_k(\varepsilon^{-\alpha}t)
	\right]
	u_{I_j} 	
	 \label{cinco}
\end{align}
Comparing the first terms of \eqref{cuatro} and  \eqref{cinco}, we
conclude that we may neglect the former; while comparing the  first term
of \eqref{cuatro} and  the third of \eqref{cinco},  that we may neglect
the latter.  Also, comparing the  first term and second terms  of
\eqref{cinco}, we see that we may neglect the latter.
We are left with:
\begin{align}
\dot u_{\theta_i} &=
	 \varepsilon^{-\alpha+\beta}
	  \sum_{ j}\frac{\partial^2 H }{\partial I_i \partial I_j}
\;	u_{I_j}  \label{seis}
\\
\dot u_{I_i} &=
	\varepsilon^{\frac12-\alpha-\beta} \sum_{j,k}
		\frac{\partial^2 G_k}{\partial \theta_i \partial \theta_j}(\varepsilon^{-\alpha}t)
			\; \xi_k(\varepsilon^{-\alpha}t)
	\;u_{\theta_j}
	 \label{siete}
\end{align}
which is understood for white noise in the Ito convention.
Here the $\frac{\partial^2 H}{\partial I_i \partial I_j}$ are constants,
and the $\frac{\partial^2 G_k}{\partial \theta_i \partial \theta_j}$
quantities that are depend on time through the angles $\theta_i(t)$,
defined by the torus and given by \eqref{hami}.

This is as far as we can go for a general perturbation. If we now we
consider the case of white noise, we have that:
$ \xi_k(\varepsilon^{-\alpha}t) = \varepsilon^{\frac \alpha 2}
\xi_k(t)$. We may proceed as follows: we choose
$\alpha=\frac 13$ and
$\beta=\frac 13$:
\begin{align}
\dot u_{\theta_i} &=
	  \sum_{ j}\frac{\partial^2 H }{\partial I_j \partial I_i}
\;	u_{Ij}  \label{aa}
\\
\dot u_{I_i} &= 
\sum_{j,k} \frac{\partial^2 G_k}{\partial \theta_i \partial \theta_j}(\varepsilon^{-\alpha}t)\;
	\xi_k(t)
	\;u_{\theta_j} 
	 \label{bb}
\end{align}
This is not yet the final product. We have to note now that the 
$\frac{\partial^2 G_k}{\partial \theta_i \partial \theta_j}(\varepsilon^{-\alpha}t)$
are rapidly oscillating functions of (rescaled) time. 

We now use the fact that in the limit of high frequency (in our case $\varepsilon \rightarrow 0$), one
 may replace the oscillating terms by their root mean square  average. 
To see that this is generically
the case, consider a stochastic process with generator ${\cal L}(\omega t)$, a periodic function of time.
The generating function over one period is given by the time-ordered exponential 
${\cal T} e^{-\int {\cal L}(\omega t) dt}=e^{- {\cal L}_{av} t}$, where the averaged generator may be developed 
using the Magnus expansion \cite{Tannor}:
\begin{equation}
{\cal L}_{av} = \frac{\omega}{2 \pi} \left[ \int_0^{\frac{2\pi}{\omega}} dt \;  {\cal L}(\omega t) dt
+\int_0^{\frac{2\pi}{\omega}} dt  \; \int_0^t dt'  [{\cal L}(\omega t), {\cal L}(\omega t')] +...\right]
\end{equation}
Rescaling times, one finds that the second term is of order $\omega^{-1}$, the subsequent one $\omega^{-2}$,  and so on.
Averaging over time the generator means, going back to the equation (\ref{bb}) which is in Langevin form,  that we substitute
the noises terms by  white, correlated Gaussian noises $\rho_{ij}$with correlations:
\begin{equation}
\begin{aligned}
\Lambda_{ijkl} &=  \frac{1}{\hat t}
 \int_0^{\hat t} dt \int_0^{\hat t'} dt'  \frac{\partial^2 G_k}{\partial \theta_i \partial \theta_j}(\varepsilon^{-\alpha}t)\; 
    \frac{\partial^2 G_k}{\partial \theta_k \partial \theta_l}(\varepsilon^{-\alpha}t')\;  \overline{\xi_k(t)
  \xi_k(t')} 
\\
 &= \frac{1}{\hat t} \int_0^{\hat t} dt \left[ \frac{\partial^2 G_k}{\partial \theta_i \partial \theta_j}
    \frac{\partial^2 G_k}{\partial \theta_k \partial \theta_l} \right](t)
\end{aligned}
\end{equation}
where $\hat t$ is a time that is long enough that it encompasses an almost integer number of cycles of the variables,
but is short with respect to $\varepsilon^{-1/3}$. We finally obtain:
\begin{align}
\dot u_{\theta_i} &=
	  \sum_{ j}\frac{\partial^2 H }{\partial I_i \partial I_j}
\;	u_{I_j}  \label{aaa}
\\
\dot u_{I_i} &=  \sum_j { \rho}_{ij}(t) u_{\theta j}
	 \label{bbba}
\end{align}
with $\overline{\rho_{ij}(t)\rho_{kl}(t')} = 2 \Lambda_{ijkl} \delta(t-t')$.  
We are now in a position to write the equation for the evolution
equation of the probability distribution $P(u_{\theta_i},u_{I_i})$
of the $u_i$:
\begin{equation}
\frac{\partial P}{\partial t}
=
\left[
\sum_{ij}
\left(\frac{\partial^2 H}{\partial I_i \partial I_j} \right) \;
	u_{I_j} \frac{\partial}{\partial u_{\theta_i}}
+
\sum_{ijlm}
\Lambda_{ijlm}
\frac{\partial^2}{\partial u_{I_i} \partial u_{I_j}}
u_{\theta_l} u_{\theta_m}
\right]
P
\label{fp}
\end{equation}
Note that in \eqref{aaa}, \eqref{bbba} and \eqref{fp} time here has been rescaled as
$t \rightarrow \varepsilon^{-1/3} t  $ (cfr Eq~\eqref{uuy}).

\section{A single degree of freedom\label{sec:1dof}}

Let us now specialize to a single degree of freedom.
The equations \eqref{aaa} and \eqref{bbba} read, in this case:
\begin{align}
\dot u_{\theta} &= \frac{d^2 H}{d I^2} \; u_{I}  \label{aaa11} \\
\dot u_{I} &= {\rho}(t) u_{\theta }              \label{bbb}
\end{align}
with $\overline{\rho(t)\rho(t')} = \delta(t-t') \Lambda_{II\theta \theta}$.
The root mean square geometric factor for the amplitude of the noise reads:  
\begin{equation}
\Lambda_{II\theta \theta}
=
\frac{1}{\hat t} \int_0^{\hat t} dt \left[\frac{\partial^2 G_k}{\partial \theta^2}(t) \right]^2
\equiv
\overline{\left(\frac{\partial^2 G}{\partial\theta^2}\right)^2}
=
\overline{\left(\frac{\ddot G}{\omega(I)^2}\right)^2}
  \end{equation}
where we have used the fact that $\frac{\partial}{\partial\theta}
= \frac{dt}{d\theta} \frac{\partial}{\partial t}
= \frac1{\omega(I)} \frac{\partial}{\partial t}$ and  $ \omega(I) \equiv \frac{dH}{dI} $.
The corresponding Fokker-Planck equation is:
\begin{equation}
\frac{\partial P}{\partial t}
=
\left[
\left(\frac{d^2 H}{d I^2} \right) \;
	u_{I} \frac{\partial}{\partial u_{\theta}}
+
\Lambda_{II \theta \theta}
\frac{\partial^2}{\partial u_{I}^2 }
u_{\theta}^2
\right]
P
\label{fp11}
\end{equation}

In the one degree of freedom case, we may now perform a further
rescaling of $t$ and $u_\theta$, and obtain an  adimensional equation for
the evolution of the probability $\tilde P$ of the rescaled variables:
\begin{equation}	\label{eq:fokkerplanck1d:rescaled}
\frac{\partial \tilde P}{\partial \bar t}
=
\left[
- \frac{\partial }{\partial \tilde u_\theta} u_I 
+ 
	\frac{\partial^2}{\partial u_I^2} {\tilde u_\theta}^2
\right]
\tilde P(\tilde u_\theta,u_I,\bar t)
.
\end{equation}
This equation appears frequently in the theory of one-dimensional
localization, and in the related problem of the harmonic oscillator with
randomly diffusing frequency (see
References~\cite{Halperin,Derrida,tessieri2000,mallick2002,schomerus2002},
whose approches we shall follow).

The rescaled time $\bar t$ is expressed, {\em with respect to the original time} $t$ as:  $\bar t= \frac t \tau$ where 
$\tau$ is the characteristic time
\begin{equation}	\label{eq:tau}
\tau = \left[
  \varepsilon \,
  \overline{\left(\frac{\partial^2 G}{\partial\theta^2}\right)^2}
  \left(\frac{d\omega}{dI}\right)^2
\right]^{-1/3}
= 
\left[
  \varepsilon \,
  \overline{(\ddot G)^2}
  \left(\frac1\omega \frac{d\omega}{dH}\right)^2
\right]^{-1/3}
.
\end{equation}
here we have used that
$\frac{d\omega}{dI}
= \frac{d\omega}{dH} \frac{dH}{dI}$.
The factor $\left(\frac1\omega \frac{d\omega}{dH}\right)^2$
appearing in the characteristic time~\eqref{eq:tau} 
is a measure of the  difference is period of neighboring orbits, and we shall hence call it
\emph{isochronicity parameter}. It is zero for a harmonic oscillator.
Denoting $t_P=2\pi/\omega$ the period of oscillations, we may also write
\begin{equation}
\left(\frac1\omega \frac{d\omega}{dH}\right)^2
=
\left(\frac1{t_P} \frac{d t_P}{dH}\right)^2
.
\end{equation}

 Starting from an initial length $\|u(0)\|=1$, we define the (quenched)  Lyapunov exponent as the
 average of the logarithmic separation:
\begin{equation}
\lambda(t) = \frac1t \langle \ln{\|u(t)\|} \rangle
.\label{quenched_lyap}
\end{equation}
An {\em annealed} estimate may be also defined as:
\begin{equation}
\lambda^{(2)}(t) = \frac1{2t} \ln \langle {\|u(t)\|^2} \rangle
.\label{annealed_lyap}
\end{equation}
where averages are taken over the stochastic noise realizations.
Because all the dependence on the problem is through the timescale $\tau$, we have that both exponents are 
proportional to $\tau^{-1}$, with different dimensionless proportionality constants of order one.  

\vspace{.5cm}

{\bf Annealed  Lyapunov exponent  $\lambda^{(2)}$}

\vspace{.5cm}

The annealed Lyapunov exponent is easy to calculate using the
property~\cite{mallick2002} that the moments of order two 
\begin{equation}	\label{eq:moment2}
\langle u_a u_b \rangle = \int u_a u_b P(u_\theta,u_I,t)
	\,
	du_\theta
	\,
	du_I
.
\end{equation}
evolve through a closed system of equations. Using equation (\ref{fp11}) one may easily see that, to leading order in
$\varepsilon$:
\begin{equation}
\frac{d}{dt}
\begin{pmatrix}
\langle u_\theta u_\theta \rangle \\
\langle u_\theta u_I \rangle \\
\langle u_I u_I \rangle \\
\end{pmatrix}
=
2
\left[
\left(\frac{d\omega}{dI}\right)
\begin{pmatrix}
0 &1 &0 \\
0 &0 &\frac12 \\
0 &0 &0 \\
\end{pmatrix}
+
\varepsilon
\begin{pmatrix}
0 & 0 & 0 \\
0 & 0 & 0 \\
\Lambda_{II\theta \theta} & 0 & 0
\end{pmatrix}
\right]
\begin{pmatrix}
\langle u_\theta u_\theta \rangle \\
\langle u_\theta u_I \rangle \\
\langle u_I u_I \rangle \\
\end{pmatrix}
\end{equation}

The largest eigenvalue $\mu_M$ of the matrix in the right hand side yields the annealed Lyapunov exponent $\lambda^{(2)}$.
 The eigenvalue equation is easy to derive:
 \begin{equation}
\mu_M^3
=
\frac12
\varepsilon
\Lambda_{II\theta \theta}
[\omega'(I)]^2 = \frac{1}{2} \tau^{-3}
\end{equation}
and we get:
\begin{equation}	\label{eq:lyap_q=2}
2 \lambda^{(2)}
= 
\mu_M = \frac{2^{-1/3}}\tau
\end{equation}
We easily check that
$\langle u_\theta u_I\rangle \propto \varepsilon^{1/3} \langle u_\theta^2\rangle$
et
$\langle u_I^2\rangle \propto \varepsilon^{2/3} \langle u_\theta^2\rangle$;
which means that the if the Lyapunov vector has a component of order one along
the $\theta$ direction (tangent to the torus), it has a component of order
$\varepsilon^{1/3}$ along the $I$ direction ({\it i.e.} transverse to the
torus).

\vspace{.5cm}

{\bf `Quenched' Lyapunov exponent}

\vspace{.5cm}

In order to have a more complete description, it is useful to introduce the
Riccati variable~\cite{Halperin,Derrida}:
\begin{equation}	\label{eq:def_z}
z = \frac{u_I}{u_\theta} = \left(\frac{d\omega}{dH}\right)^{-1} \left(\frac{\dot u_\theta}{u_\theta}\right) 
\end{equation}
Clearly, the average Lyapunov exponent is given by:
\begin{equation}
\lambda = \omega' \langle z \rangle =\langle \frac{\dot u_\theta}{u_\theta}\rangle
\end{equation}

When $z$ is  introduced in the Ito version of the
Langevin~(\ref{aaa}) and (\ref{bbb}) we get:
\begin{equation}	\label{eq:langevin_z}
(I)
\qquad
\dot z = -\omega'(I) z^2 + \varepsilon^{1/2} \xi(t)
.
\end{equation}
with $\overline{\xi(t) \xi(t') } = \Lambda_{II \theta \theta} \delta(t-t')$ 
From this, or directly from (\ref{fp11}), we obtain the Fokker-Planck version:
\begin{equation}
\frac{\partial P}{\partial t}
=
\left[
\omega'(I) \frac{\partial}{\partial z} z^2
+
\varepsilon
\Lambda_{II \theta \theta} 
\frac{\partial^2}{\partial z^2}
\right]
P(z,t)
.
\label{fpz}
\end{equation}
Again,  we may rescale out all physical constants:
\begin{align}
\tilde t &= t / \tau \\
\tilde z &= z / h
\end{align}
where $\tau$ given in~\eqref{eq:tau}
and
\begin{equation}	\label{eq:h}
h
= \bigl[\tau \omega'(I)\bigr]^{-1}
= \Bigl(\tau \omega\frac{d\omega}{dH}\Bigr)^{-1}
.
\end{equation}
We get:
\begin{equation}	\label{eq:fokkerplanck:tildez}
\frac{\partial \tilde P}{\partial \tilde t}
=
\frac{\partial}{\partial \tilde z}
\left[
\tilde z^2
+
\frac{\partial}{\partial \tilde z}
\right]
\tilde P(\tilde z,\tilde t)
\end{equation}
(which we could have obtained directly from (\ref{eq:fokkerplanck1d:rescaled})), and:
\begin{equation} \label{eq:langevin:tildez}
(I)\qquad
\qquad
\dot {\tilde z} = -\omega'(I) \tilde z^2  + \tilde \xi(t)
\end{equation}
where $\tilde \xi(t)$ is a Gaussian white noise of  variance 2.
%

%

In order to calculate the Lyapunov exponent, we need the expectation value of $\bar z$, computed with the stationary solution 
of Equation~\eqref{eq:fokkerplanck:tildez}
satisfying
$\partial \tilde P_\infty/\partial\tilde t=0$. 
Note that we are trying to solve for the stationary solution of a particle in an unbounded (cubic) potential. This is in fact
impossible unless we re-inject at $-\infty$ particles that have reached $+\infty$: the stationary state has a constant current. 
This is not as strange as it seems, because as we shall see below, $z$ has the interpretation of the tangent of an angle which grows monotonically.
The solution we find is then:
\begin{equation}
\tilde P_\infty(\tilde z)
=
  \frac1{\cal N} \exp\left(-\frac{\tilde z^3}{3}\right)
    \left[ C  +
     \int_0^{\tilde z}
     	\exp\left(\frac{\tilde y^3}{3}\right)  \,
     d \tilde y\right]
 = \frac1{\cal N}
    \int_{-\infty}^{\tilde z} 
    	\exp\left(\frac{{\tilde y^3 - \tilde z^3}}3\right)
\,    d \tilde y 
\end{equation}
where we have put
\begin{equation}
C = \int_{-\infty}^0 \exp\left(\frac{\tilde y^3}{3}\right) \, d \tilde y 
\end{equation}
in order to assure normalizability and positivity.
The normalization constant is given by~\cite{schomerus2002}:
\begin{equation}
{\cal N}
=
\int_{-\infty}^{\infty} 
    \int_{-\infty}^{\tilde z} 
    	\exp\left(\frac{{\tilde y^3 - \tilde z^3}}3\right)
\,    d \tilde y 
\,    d \tilde z
=
\pi^2 \left[ \text{Ai}^2(0) + \text{Bi}^2(0) \right]
=
\left(\frac23\right)^{1/3} \frac{\sqrt{\pi} \, \Gamma(\frac16)}{\sqrt 3}
\simeq
4.97605
\end{equation}
where $\text{Ai}$ et $\text{Bi}$ are the Airy functions.
The average  $\langle \tilde z \rangle $ is readily obtained as:
\begin{equation}	\label{eq:avg_tildez}
\langle \tilde z \rangle= \int d \tilde z\;  \tilde P_\infty(\tilde z) \, \tilde z
=
\left(\frac32\right)^{1/3} \frac{\sqrt\pi}{\Gamma(\frac16)}
\simeq
0.364506.
\end{equation}
The Lyapunov exponent is then:
\begin{equation}	\label{eq:lyap_quenched0}
\lambda
= \omega'(I) \langle z \rangle
= \frac{\langle\tilde z\rangle}{\tau}
=
\left(\frac32\right)^{1/3} \frac{\sqrt\pi}{\Gamma(\frac16)}
\left[
  \varepsilon \,
  \overline{(\ddot G)^2}
  \left(\frac1\omega \frac{d\omega}{dH}\right)^2
\right]^{1/3}
.
\end{equation}

\begin{figure}
\begin{center}
\includegraphics[angle=90]{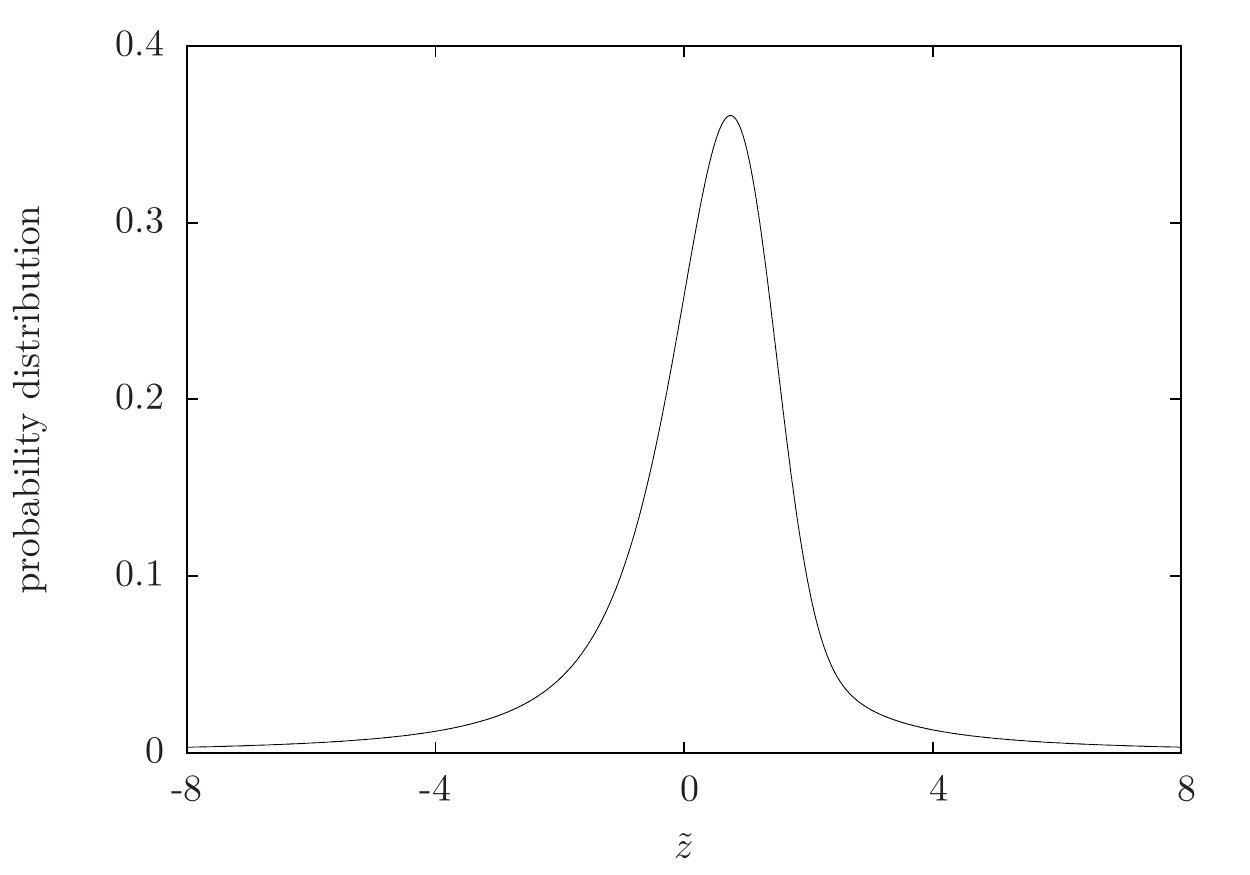}
\end{center}
\caption{	\label{fig:pdf_z}
Stationary distribution of $\tilde z$.
}
\end{figure}

\section{Lyapunov jumps and Lyapunov-vector phase-slips}

\vspace{.5cm}

{\bf Evolution of the direction of the Lyapunov vector}

\vspace{.5cm}

\begin{figure}
\begin{center}
\includegraphics{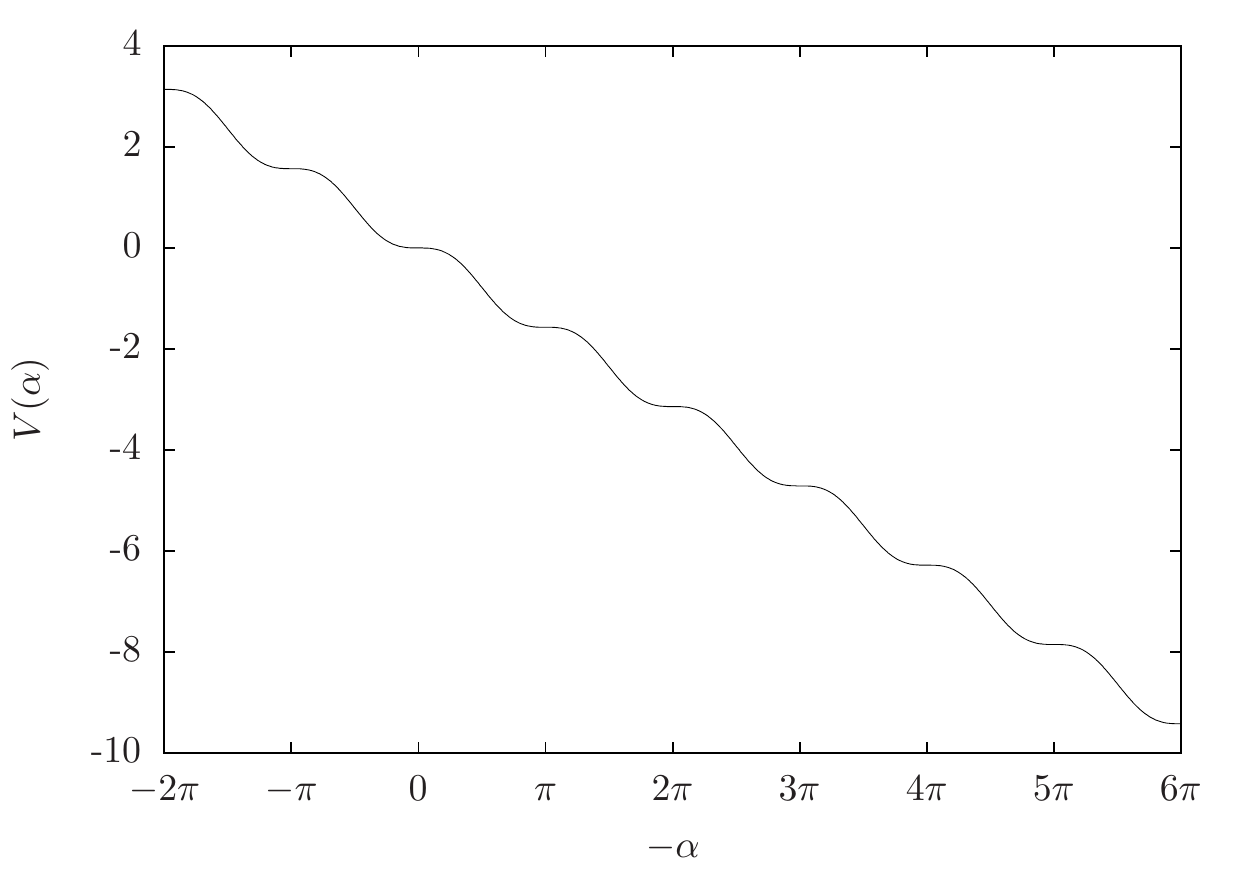}
\end{center}
\caption{	\label{washboard}
Potential for the angle  $\alpha$.
}
\end{figure}

Let us introduce the angle of the Lyapunov vector as:
\begin{equation}
\alpha = \arctan z = \arctan\left({\frac{u_I}{u_\theta}}\right)
\end{equation}
\begin{figure}
\begin{center}
\includegraphics{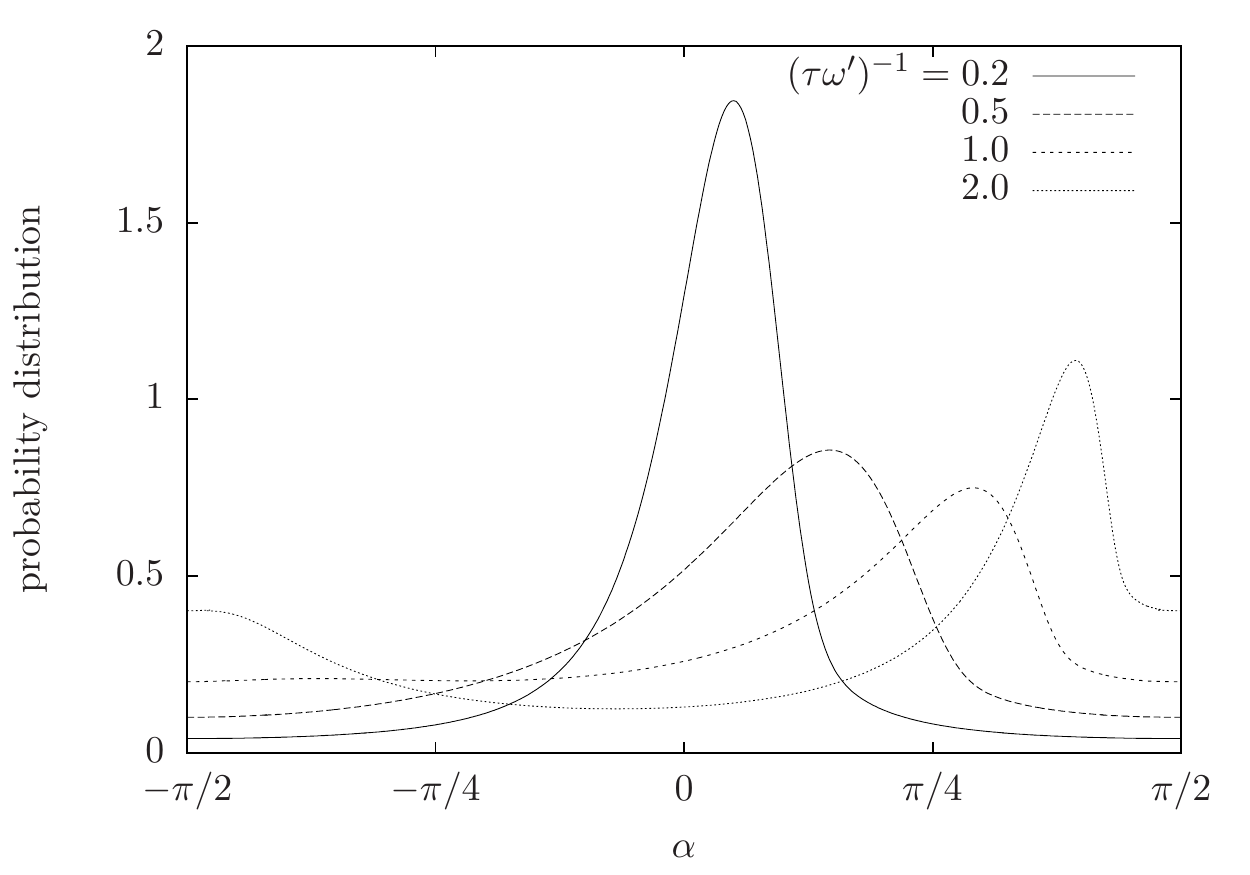}
\end{center}
\caption{
The stationary distribution of the angle $\alpha$ for different values of parameters.
}
\end{figure}
The evolution of the probability distribution $P_\alpha(\alpha,t)$ may be
obtained directly by changing variables in the Fokker-Planck
equation~\eqref{fpz}, to get:
\begin{equation}
\frac{\partial P_\alpha}{\partial  t}
=
\frac{\partial}{\partial \alpha} \left[
    \frac{dV}{d\alpha}	
+
    (\tau \omega')^{-2}
    \frac{\partial}{\partial \alpha} \cos^4\alpha
\right] P_\alpha
\end{equation}
where 
\begin{equation}
V(\alpha) = -\frac{\omega'}2 \left(
	\alpha
	- \frac12 \sin{2\alpha}
	- (\tau \omega')^{-3} \cos^4\alpha
\right)
.
\end{equation}
Equivalently, we  find that the angle $\alpha$ follows a Langevin
equation:
\begin{equation}
(I)\qquad
\dot \alpha
=
-\frac{dV}{d\alpha}
+
(\tau \omega')^{-1} \cos^2\alpha  \; \xi(t) \approx -\frac{dV}{d\alpha}
+
(\tau \omega')^{-1}  \; \xi(t)
\end{equation}
where we have replaced  the cosine by one, the value it takes at the
only times when the noise is non-negligible.  The system  is a {\em
marginal washboard} potential (Fig. \ref{washboard})  with very small
corrections and small noise.
Away from saddles, the angle evolves {\em monotonically} and almost
deterministically: these are the `phase slips'. This deterministic
motion  by itself would leave it trapped in the saddles: here the effect
of noise -- or in general any form of perturbation -- is crucial,
because it allows the system to traverse the saddle and start a new
phase slip.
Because the noise is weak, most of the time is spent around saddles
where $\alpha=0 \mod 2\pi$, and for those times  the Lyapunov vector
stays tangent to the torus.

Let us see what happens during a phase slip. During those times, we may
neglect the noise in the equation for $z$. Solving the deterministic
equation $\dot{\tilde z}=-{\tilde z}^2$ with some initial condition
$u_\theta(0),u_I(0)$ we obtain: $u_I(t) = -|u_I(0)|$   and $u_\theta(t)
= |u_\theta(0)| - |u_I(0)| t$.
The norm of the vector evolves smoothly until the slip starts, then dips
to a minimum of $\sqrt{u_I(t)^2 + u_\theta(t)^2} \sim |u_I(0)|$  which
is achieved at half-slip $\alpha=\frac \pi 2$, and then quickly recovers
in the next half-slip what it had lost during the first.

The average time elapsed between phase slips is proportional to the
Lyapunov time. In order to compute this we calculate the flux of $z$
defined from the Fokker Planck equation
$\partial\tilde P/\partial\tilde t = -\partial\tilde j/\partial \tilde z$
at stationarity. 
\begin{equation}
\tilde j(\tilde z) = - \frac{1}{{\cal N}\tau}
.
\end{equation}
The average time between slips is simply given by $1/|j|$:
\begin{equation}
\langle t_{slip} \rangle
= \frac{1}{|j|}
= {\cal{N} \tau}
= \frac{\pi}{\sqrt 3} \, \lambda^{-1}
\simeq
1.8138 \, \lambda^{-1}
.
\end{equation}
Let us see how this comes about in a simple example, the dynamics with $\omega'=\Lambda_{II \theta \theta}=1$.
\begin{equation}	\label{eq:sysideal}
(I)\qquad
\begin{aligned}
\dot u_\theta &= u_I \\
\dot u_I &= \varepsilon^{1/2}  \xi(t) \; u_\theta
\end{aligned}
\end{equation}
and  $\varepsilon=10^{-3}$.
Equations are of the form  \eqref{aaa11} and \eqref{bbb}.
We start with a random vector $u(0)$ with unit nor  $\|u(0)\|=1$ 
and random orientation $\alpha(t=0)$.

\begin{figure}
\begin{center}
\includegraphics{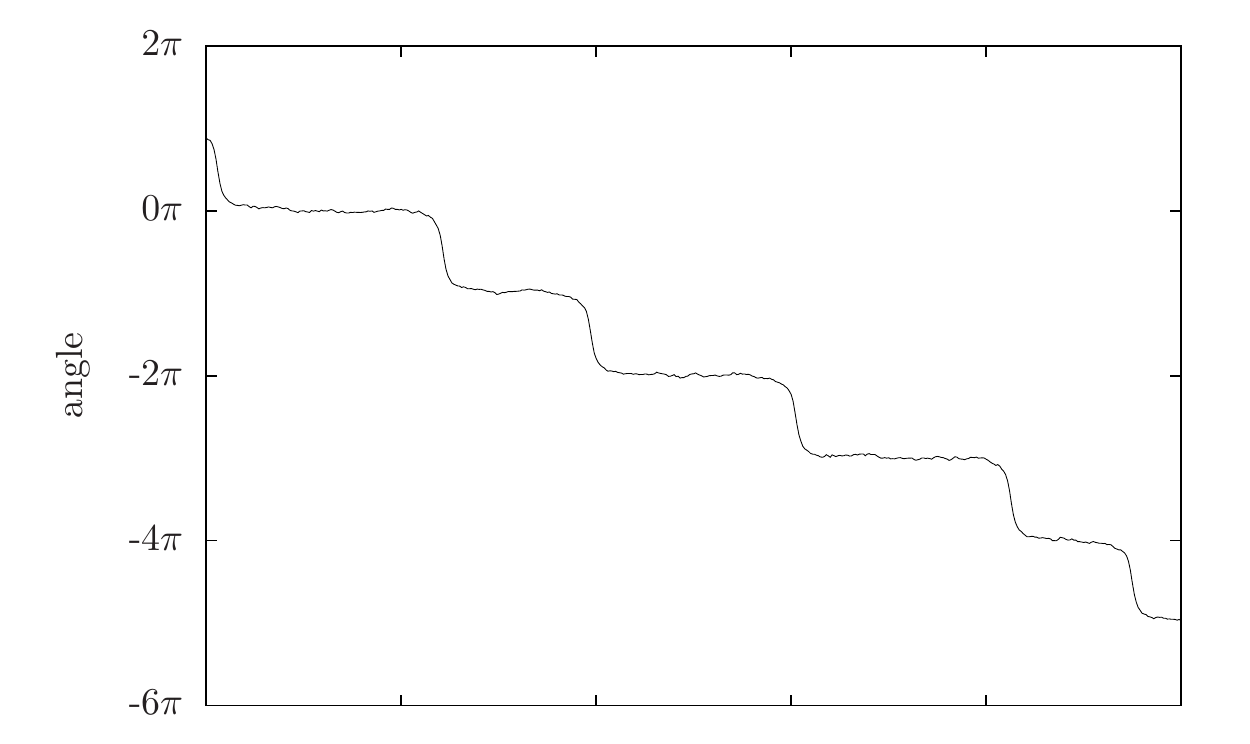}\\
\includegraphics{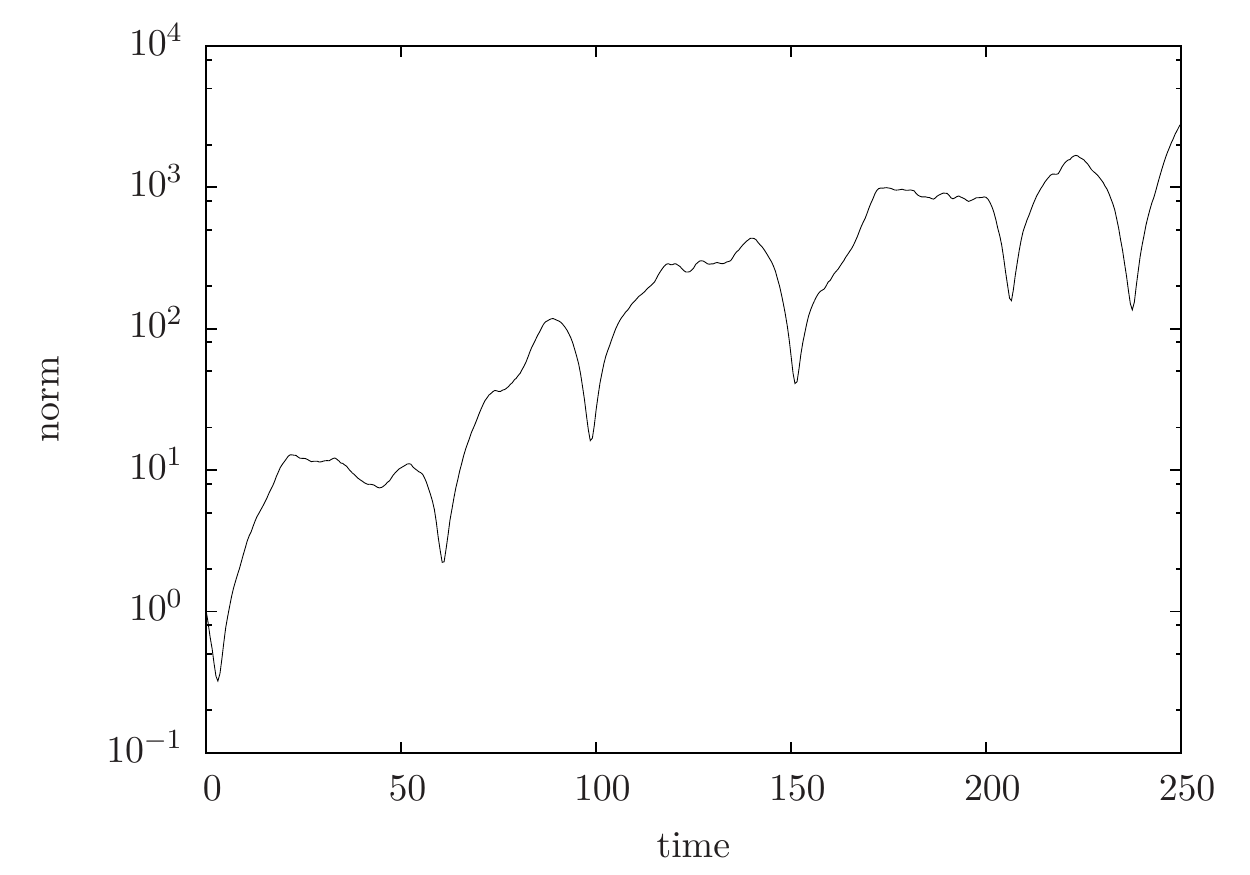}
\end{center}
\caption{	\label{fig:traj_sys1}
Evolution of the Lyapunov angle $\alpha$  and norm for  $\varepsilon=10^{-3}$.
The characteristic time is $\tau=10$.
}
\end{figure}

Figure~\ref{fig:traj_sys1} shows the evolution of $\alpha$ (which should
be considered only modulo $2 \pi$): the phase slips are clearly visible.
Whenever there is phase-slip, the finite time Lyapunov exponent shows a
dip.  These general features are clearly visible  in the computations of Lyapunov exponents of
planetary motion~\cite{sussmanwisdom}.
Although the Lyapunov vector  is unfortunately not generally quoted in
those cases, one expects that phase slips are the cause of the dips also
for planets.

\section{The role of separatrices: the example of the simple pendulum \label{sepa}}

\begin{figure}
\begin{center}
\includegraphics{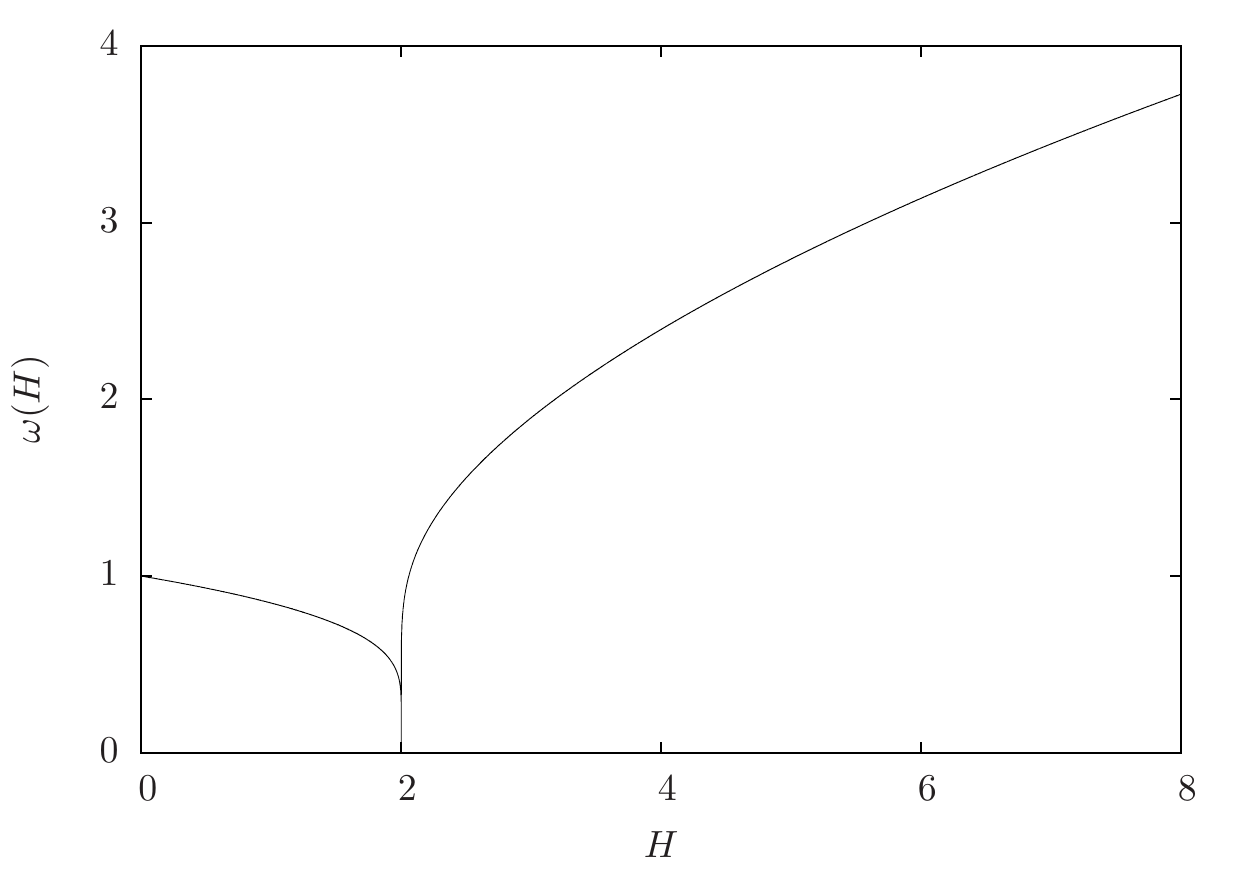}
\end{center}
\caption{	\label{pendu}
Frequency versus energy for the simple pendulum
}
\end{figure}

As one would expect, the instability of trajectories is larger in and
around separatrices.  In order to see this, consider the example of the
simple pendulum $H=\frac12 p^2 + 1-\cos{q}$.
The frequency in terms of the energy is shown in Fig~\ref{pendu}.

Small oscillations correspond to the linear regime, for which one
has~\cite{pendu_p}
$\left.\frac{dT}{dH}\right|_{H=0}=\frac{T_o}{8mg}  \neq 0$
so that even a small amplitude trajectory will develop an instability in
the presence of noise.  The neighborhood of a separatrix  $H=2$ is also
interesting.
For $\delta \equiv |H - 2|$, one may compute
\begin{equation}
\omega(\delta \to 0) \simeq \frac{\pi}{|\!\ln\delta |} \to 0
\end{equation}
from which
\begin{equation}
\frac1\omega \frac{d\omega}{dH} \sim \frac{1}{\delta \,|\!\ln{\delta}|}
\to
\infty
.
\end{equation}
Because $\overline{(\ddot G)^2} = \overline{(\ddot q)^2}$ is of order
$\omega^4$ we find that the Lyapunov exponent scales as:
$\varepsilon^{1/3} |\!\ln{\delta} |^2 \delta^{-2/3} \to \infty$, which
means that just on the separatrix  it scales differently with $\varepsilon$.

The behavior of the pendulum is quite generic for nondegenerate fixed points.
Consider the dynamics around a fixed point, say
 $(q_0,p_0)=(0,0)$. We may assume that  $H_0=H(q_0,p_0)=0$, 
and generically to lowest order $H$ reads:
\begin{equation}
H(q,p) = \frac{p^2}{2m} + \alpha q p + \frac{k}{2} q^2
\end{equation}
where the constants  $m$, $k$, $\alpha$  may have any magnitude or sign.
Hamilton's equations read:
\begin{equation}
\ddot q + \left(\frac{k}{m} - \alpha^2\right) q = 0
.
\end{equation}
There are two possible cases: either
\begin{itemize}
\item
$\frac{k}{m} - \alpha^2 > 0$.
 the system is locally a harmonic oscillator
 $\omega_0^2 = \frac{k}{m} - \alpha^2$,
and the fixed point is elliptic.
The development of $\omega$ is to first order
$\omega = \omega_0 + \Delta\omega(H-H_0)$
so that $\left(\frac{d\omega}{dH}\right)^2$ is a minimum at the center of the ellipse. 
For a fixed value of the  noise, in this point the  Lyapunov exponent is minimal.
%
\item
$\frac{k}{m} - \alpha^2 < 0$.
The fixed point is hyperbolic, the trajectory is a separatrix.
Consider a trajectory starting close to this point, of energy  $H=\delta\ll 1$.
The dynamics in  $(q,p)$ starts along the unstable direction and the distance
grows as 
 $\sim \sqrt\delta e^{\omega_0 t}$
où $\omega_0 = |\frac{k}{m} - \alpha^2|$, 
becoming of order one 
$\sqrt\delta  e^{\omega_0 t} \sim 1$,
after a time of order $\frac1{\omega_0} |\!\ln\delta|\gg 1$.
Once the system is away from the critical point, its subsequent evolution takes
a time of order one.  We hence conclude that the frequency close to an elliptic
points goes as:
\begin{equation}
\omega \sim \frac{\omega_0}{|\!\ln\delta|} \ll 1
.
\end{equation}
\end{itemize}

We thus find that the behavior near minima and separatrix of the pendulum is
generic for nondegenerate ($\omega_0 \neq 0$) situations.

\section{Analogies}

In this section we discuss two illuminating analogies that give us a better intuitive understanding of the phenomenon 
we discuss in this paper.

\subsection{Polymer tumbling in a laminar flow}

A polymer in a flowing liquid tends to align with the direction of flow. 
If we consider that the fluid is at finite temperature, thermal fluctuations will make the polymer 
misalign with the flow. Now, if there is a local share rate, the speed at one end of the polymer will be higher, and at the other end lower,
than that of its center of mass, ultimately forcing it to tumble through a half-turn ~\cite{misha}
Clearly, the tumbling frequency goes to zero in the low-temperature limit in which the noise amplitude is negligible.
The phenomenon is closely analogous the the slips of the Lyapunov vector. This analogy does not extend to the actual length of
the vector itself, because the ends of the polymer are not free to diverge, but are kept at finite distance by the elasticity.

\subsection{Anderson localization}

Let us discuss the close physical analogy  between our problem and
Anderson Localization.It will become clear that the situation we are dealing with is critical: in
our language it is in the limit between a regime with exponentially rare (in
terms of  $\varepsilon$) phase slips, and a regime with frequent ($O(1)$)
slips. This criticality shows up in the  localization language in that the
system corresponds to a {\em band edge.}

Consider equations \eqref{aa} and \eqref{bb} and eliminate the $u_{I_i}$. We
get, in the Ito convention:
\begin{equation}
\ddot u_{\theta_i} + \sum_j \left\{-
	\sum_{k l}\frac{\partial^2 H }{\partial I_i \partial I_l  }
	     \frac{\partial^2 G_k}{\partial \theta_j \partial \theta_l}(\varepsilon^{-\alpha}t)\; \xi_k(t) \right\} \;
	 u_{\theta_j} = \ddot u_{\theta_i} + \sum_j  {\hat H}_{ij} u_{\theta_j}
 \label{ab1}
\end{equation}
which defines $\hat H_{ij}(t)$ as the term in  brackets. 
If we now make the correspondence $u_{\theta i} \rightarrow \psi_i$ and $t
\rightarrow x$, we may write the Shroedinger eigenvalue equation
\begin{equation}
\nabla^2 {\bf \psi} + {\bf \hat H \psi} = e {\bf \psi}
\end{equation}
where $\psi$ is an $N$ component wavefunction of $x$.
Our problem concerns what happens around `energy' $e=0$. 
Lyapunov exponents are related to the decay of $\psi$ for large $x$, and this
is indeed a question of localization of wavefunctions in the presence of a
potential $\hat H$. This relation has been long understood, and indeed we have
used several results originally thought for localization problems (cfr refs.
\cite{Halperin,Derrida}).

Consider the problem in one dimension. 
The Lyapunov exponent is related to the exponent in the decay of a localized
function. On the other hand, phase slips are related to the {\em nodes} in the
eigenfunction. The number of nodes of the $k$-th  wavefunction of a
one-dimensional problem is precisely $=k$ \cite{Landau}. We conclude that the
number of slips per unit length ({\it i.e.} per unit time in our original problem)
is equal to the integral of the density of levels below $e$ \cite{Derrida}.  If
there are on average no levels below, we are outside the band and there are
exponetially few  phase slips (in terms of $\varepsilon$), if we are within the
band, the number of nodes per unit length is of order $\varepsilon^0$. Our case
is precisely marginal, the system is at the band's edge and the density of
phase slips is power law $\epsilon^{1/3}$.

In conclusion, we should  emphasize two points:
\begin{itemize}
\item Our problem is a one dimensional (the time) localization situation, in
the presence of weak noise. It is hence marginal, and we are in  a {\em band
edge} situation.   
\item Our potential is random if the perturbation is random, but we may still
think of cases for which the perturbation is deterministic: the problem of a
planet perturbed by the small  interaction with others is the classical
example. In the language of localization one may ask the question as to whether
a deterministic (but complicated) potential might or not be represented by a
random one.  This has a long tradition in solid state physics:  although there
are no definite universal answers, such identification gave useful insights,
perhaps the most spectacular being the explanation of Fishman, Grempel and
Prange \cite{FGP} of energy localization in `kicked' quantum systems    in
terms of Anderson localization.
\end{itemize}

\section{More general types of perturbation}

\subsection{Non Gaussian noise}

One expects that any  Markovian noise with a non-Gaussian distribution
will give the same results as the Gaussian with the corresponding
variance. the reason is that the noise is weak, so what matters is 
its cumulative effect over time, and this is in fact Gaussian by a
central limit theorem   property.  Formally, this may be seen at the
level of equation (\ref{eq:langevin_z}), writing it in
Martin-Siggia-Rose form:
\begin{align*}	\label{eq:langevin_z_1}
1 &= \int D[\xi] {\cal P}[\xi] \; \int D[z] \;  \delta\left[\dot z
       + \omega'(I) z^2 - \varepsilon^{1/2} \xi(t)\right] \\
  &= \int D[\xi] {\cal P}[\xi] \; \int D[z] \; \int D[\hat z] \;   \exp
  \left\{ i \int dt \, \hat z \left[\dot z +\omega'(I) z^2
       - \varepsilon^{1/2} \xi(t)\right] \right\} \\
  &= \int D[z] \; \int D[\hat z] \; \exp \left\{ i
          \int dt \, \hat z
            \left[\dot z +\omega'(I) z^2\right] - {\cal F}(\varepsilon^{1/2} \hat z)
      \right\}
\end{align*}
where we have introduced the noise probability $\cal P$ and the corresponding
cumulant generator $e^{{\cal F}[v]}= \int D[\xi] {\cal P}[\xi]   \exp \{ - i
\int dt \, \hat z v(t) \}$. Expanding ${\cal{F}}[\varepsilon^{1/2} ]$ in
powers of $\varepsilon$, to second order, we recover a Gaussian case.


\subsection{Noise with long time correlations}

\paragraph{Ornstein--Uhlenbeck process}

A simple way of introducing long range correlations is to consider  $\zeta(t)$ evolving as:
\begin{equation}
(I)\qquad
\dot\zeta = -\frac{1}{\tau_*} \zeta + \frac{1}{\tau_*} \xi(t)
\end{equation}
where  $\tau_*$ is the time scale of the process and $\xi(t)$ is a white noise with $\langle\xi(t)\rangle = 0$ and
$\langle\xi(t)\xi(t')\rangle = 2\varepsilon\delta(t-t')$.
This is an Ornstein--Uhlenbeck process~\cite{gardiner,risken}.
The autocorrelation reads
\begin{equation}	\label{eq:OU_corr}
\langle \zeta(t) \zeta(t') \rangle = \frac{\varepsilon}{\tau_*} e^{|t-t'| / \tau_*}
.
\end{equation}
In particular, $\langle \zeta(t)^2 \rangle =
\frac{\varepsilon}{\tau_*}$.
In the limit  $\tau_*\to 0$, $\zeta(t)$ becomes delta-correlated.

Because we now consider noise that is not white, we are not always justified  in replacing $\frac{\partial^2 G}{\partial\theta^2}$
by its root mean squared value as we did in above.
Let us write an equation for the Riccati variable without averaging over the angle variables:
\begin{equation}	\label{eq:langevin_z_zeta}
(I) \qquad
\begin{aligned}
\dot z &= -\omega'(I) z^2 + \frac{\partial^2 G}{\partial\theta^2} \zeta\\
\dot \zeta &=  -\frac{1}{\tau_*} \zeta + \frac{1}{\tau_*} \xi(t)
\end{aligned}
\end{equation}
In the limit  $\tau_* = 0$, the Lyapunov exponent
$\lambda_0$,  given by \eqref{eq:lyap_quenched0},
 inversely proportional to  $\tau_0$ given by 
\eqref{eq:tau}.
If  $\tau_*\neq 0$, $\tau_*$  is a  time scale in the problem,  
 in addition to $2\pi/\omega(I)$ and $\tau_0$.
We expect that the Lyapunov exponent $\lambda$ is given by a form
\begin{equation}
\frac{\lambda}{\lambda_0} = f\left(\lambda_0\tau_*, \omega\tau_*, \varepsilon\right)
\end{equation}
where  $f$ is adimensional.
Let shall analyze the case in  which 
$\frac{\partial^2 G}{\partial\theta^2}$ has a nonzero and a zero time average, respectively.

\paragraph{Non-zero time average}
We assume that 
$\frac{\partial^2 G}{\partial\theta^2}= \overline{\frac{\partial^2
G}{\partial\theta^2}} + {(\text{zero average term})}$, 
and  for definiteness that the time average is positive.

\begin{figure}
\begin{center}
\includegraphics[angle=90]{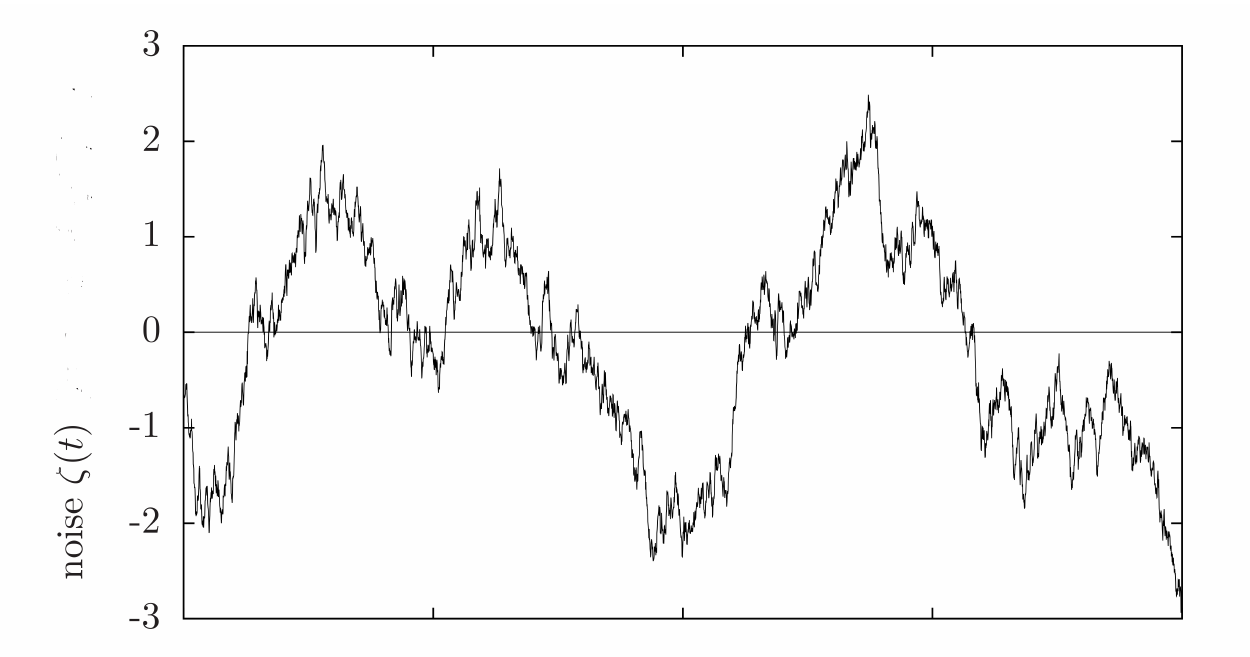}\\
\includegraphics{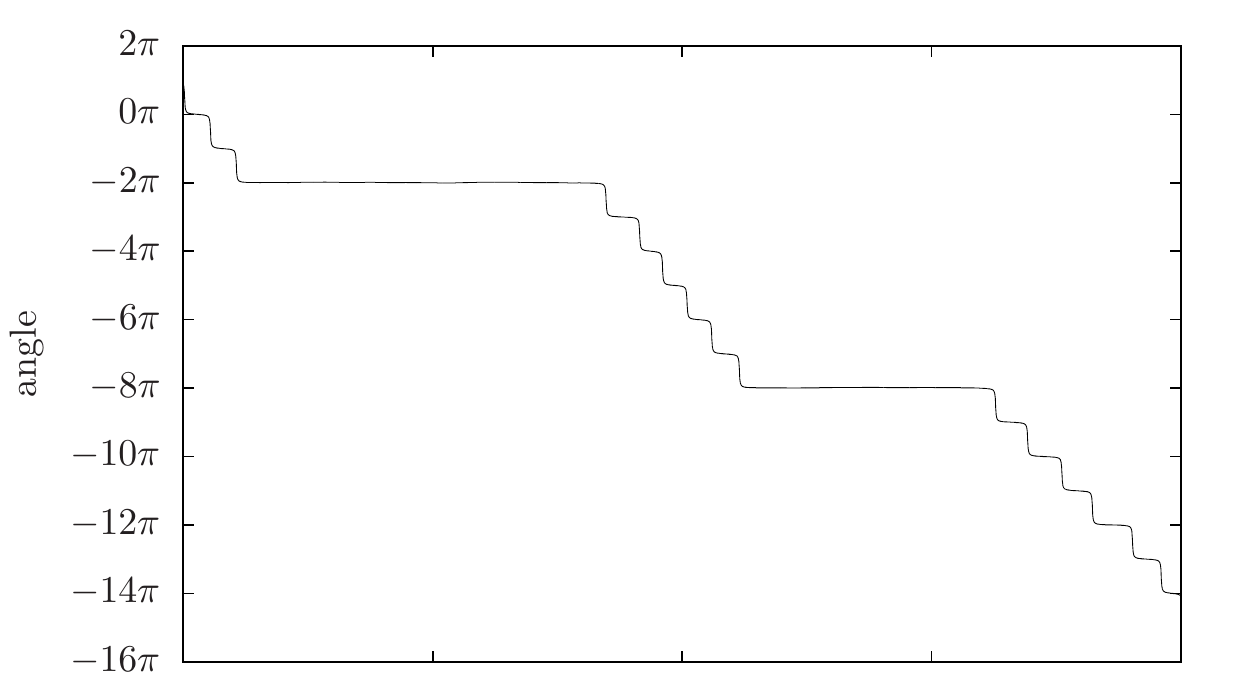}\\
\includegraphics{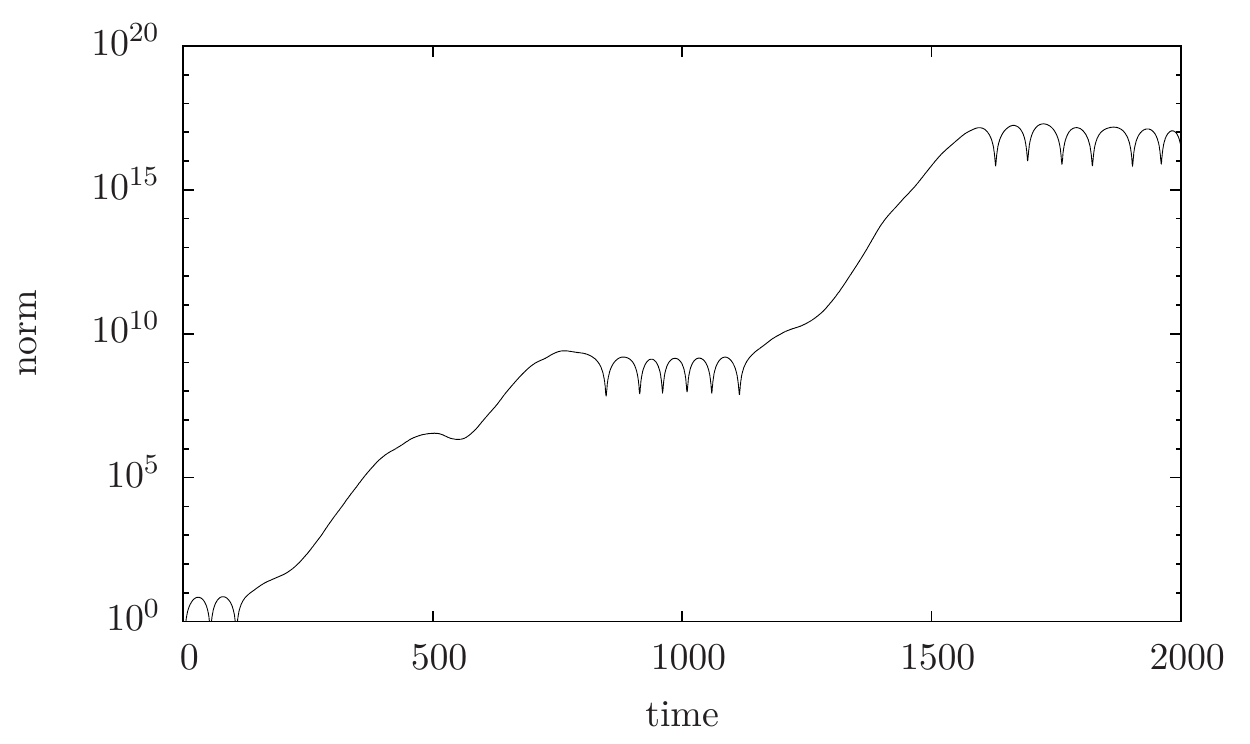}
\end{center}
\caption{	\label{fig:trajsys1corr}
A realization of the dynamics~\eqref{eq:langevin_z_zeta} with
$\varepsilon=10^{-3}$ and $\tau_*=200$,
in the case $\overline{\frac{\partial^2 G}{\partial\theta^2}}>0$.
}
\end{figure}

Assume first that  $\lambda_0\tau_*\gg 1$.
In this case, $z(t)$ is a fast variable with respect to $\zeta$.  During
the times when
\begin{itemize}
\item
$\zeta(t) < 0$.
$\dot z<0$ and the tangent vector turns rapidly.
Note that  $\zeta(t)$ stays negative for times longer than many `phase
slips'.  These events may be seen as short steps in
figure~\ref{fig:trajsys1corr}.  During such times, $z(t)$ is zero on
average, and there is no contribution to the Lyapunov exponent.
\item
$\zeta(t)>0$. During such periods:
$\zeta \sim \int_0^\infty \zeta p(\zeta) d\zeta \sim \sqrt{\varepsilon/\tau_*}$.
and $z$ follows adiabatically the equilibrium configuration for each
$\zeta$, {\it i.e.}
$\dot z \sim 0$ in \eqref{eq:langevin_z_zeta}, so that
\begin{equation}
z(t)^2
\sim
\frac{1}{\omega'(I)} \frac{\partial^2 G}{\partial \theta^2} \zeta(t)
\sim
\frac{1}{\omega'(I)} \frac{\partial^2 G}{\partial \theta^2} \sqrt{\frac{\varepsilon}{\tau_*}}
\end{equation}
from which one obtains the typical value of  $z(t)$
during such times.
\end{itemize}
The two regimes are equally probable, so that
\begin{equation}
\langle z \rangle
=
\frac12 \langle z|_{\zeta<0} \rangle
+
\frac12 \langle z|_{\zeta>0} \rangle
\sim
\left[
\frac{1}{\omega'(I)^2}
\left(
\frac{\partial^2 G}{\partial \theta^2}
\right)^{2}
\frac{\varepsilon}{\tau_*}\right]^{1/4}
\end{equation}
and we obtain
\begin{equation}	\label{eq:lyap_G2cte}
\lambda
= \omega'(I) \langle z\rangle
\sim
\left[
\frac{\varepsilon}{\tau_*}
{\omega'(I)}^2
\left(\frac{\partial^2 G}{\partial \theta^2} \right)^2
\right]^{1/4}
= \lambda_0 (\lambda_0\tau_*)^{-1/4}
.
\end{equation}

\begin{figure}
\begin{center}
\includegraphics{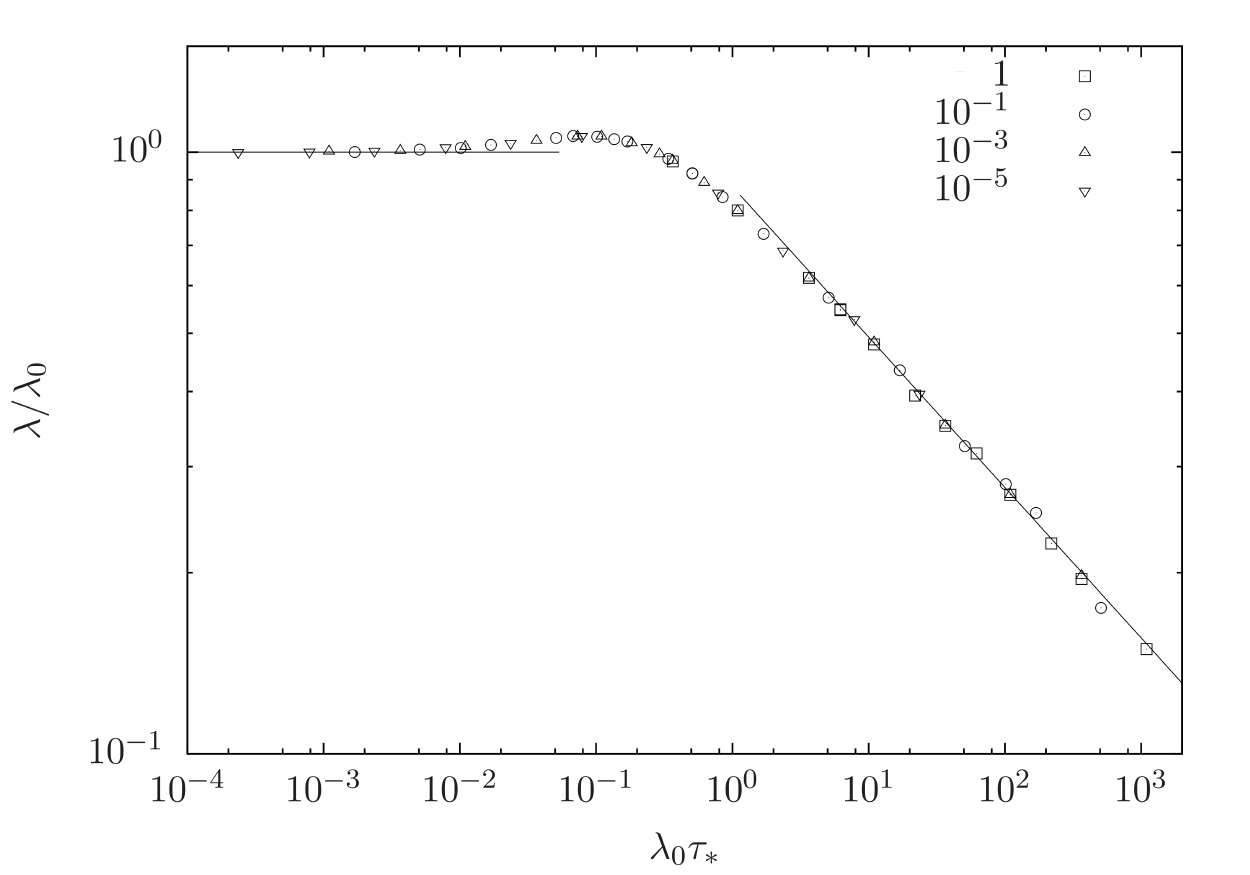}
\end{center}
\caption{	\label{fig:lyap_taustar_G2cte}
Lyapunov exponent  $\lambda$  in terms of the correlation time $\tau_*$, both adimensionalized by  $\lambda_0$.
 $\overline{\frac{\partial^2 G}{\partial\theta^2}}>0$, for  $\varepsilon=1,10^{-1},10^{-3},10^{-5}$.
The straight  lines correspond to exponents  $0$ and $-1/4$.
}
\end{figure}

Figure  \ref{fig:lyap_taustar_G2cte} shows the values of  Lyapunov exponents in terms of the parameters.
We have $\lambda/\lambda_0 \to 1$ for the regime
$\lambda_0\tau_* \ll 1$ (just as the white noise case), and
$\lambda/\lambda_0 \propto (\lambda_0\tau_*)^{-1/4}$  for the case $\lambda_0\tau_* \gg 1$.
All the dependence in  $\varepsilon$ is through $\lambda_0$.

\paragraph{Zero average: $\overline{\frac{\partial^2 G}{\partial\theta^2}}=0$ }
Here  $\frac{\partial^2 G}{\partial\theta^2}$
oscillates   with frequency $\omega$
\begin{equation}
\frac{\partial^2 G}{\partial\theta^2}
=
\Gamma \cos{\omega t}
.
\end{equation}
Clearly, in this case 
$\frac{\partial^2 G}{\partial\theta^2}$
changes sign periodically over a short timescale
$2\pi/\omega$.
and we cannot apply the arguments above.

Let us consider
$\lambda_0 \tau_* \gg 1$
and
$\omega \tau_* \gg 1$.
Because of timescale separation, we may consider instead of $\dot z$, $z^2$ and $\frac{\partial^2 G}{\partial\theta^2} \zeta(t)$,
their averages over $2\pi/\omega$,
which we shall denote:
$\overline{\dot z}$, $\overline{z^2}$ et
$\overline{\frac{\partial^2 G}{\partial\theta^2} \zeta(t)}$.
In particular,
$\overline{\cos{\omega t} \zeta(t)}$ is typically of the oder of the variation of  $\zeta(t)$ in a short period, $\overline{\dot \zeta} / \omega$.
We may thus make the same  argument as before, but considering, instead of the sugn of  $\zeta(t)$,
 the sign of  $\overline{\dot\zeta}$.

In the regime $\overline{\dot\zeta} > 0$,  the slow variable  $\overline{\dot z} \sim 0$ equilibrates, 
so that Equation~\eqref{eq:langevin_z_zeta}  averaged over time reads:
\begin{equation}
0 \sim -\omega'(I) \overline{z^2} + \Gamma \overline{\cos{\omega t}\zeta(t)}
\end{equation}
from which:
\begin{equation}
\overline{z^2}
\sim \frac{\Gamma}{\omega'(I)} \frac{\overline{\dot\zeta}}{\omega}
\sim \frac{\Gamma}{\omega'(I)} \frac{1}{\omega\tau_*} \sqrt{\frac{\varepsilon}{\tau_*}}
\end{equation}

Again, the two regimes $\overline{\dot \zeta}<0$
and  $\overline{\dot \zeta}>0$ are equiporbable, and
\begin{equation}	\label{eq:lyap_G2osc}
\lambda
= \omega'(I) \langle z\rangle
\sim
\omega'(I) {\overline{z^2}}^{1/2}
\sim
\left(
\frac{\varepsilon}{\tau_*}
\Gamma^2
\omega'(I)^2
\right)^{1/4}
(\omega \tau_*)^{-1/2}
=
\lambda_0 (\lambda_0\tau_*)^{-1/4} (\omega \tau_*)^{-1/2}
.
\end{equation}

\begin{figure}
\begin{center}
\includegraphics{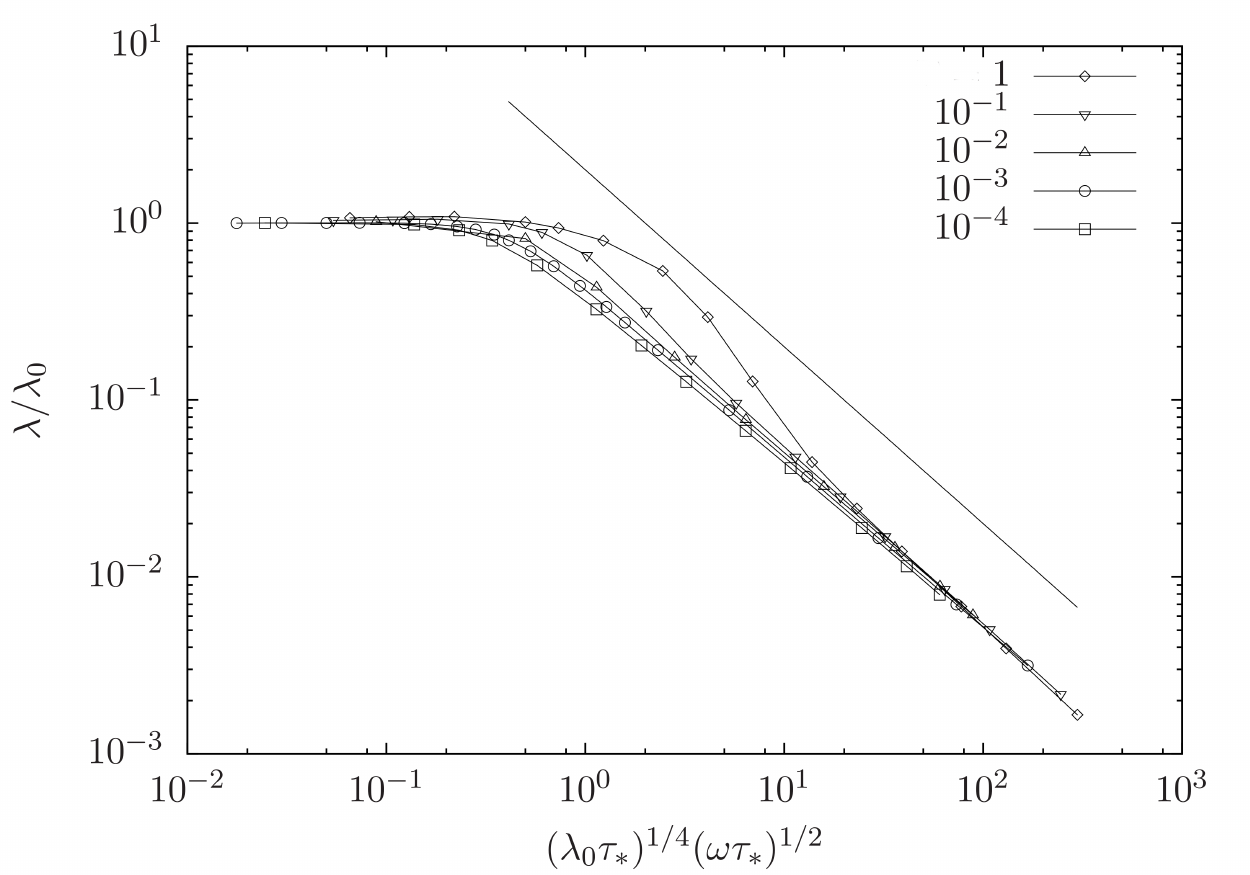}
\end{center}
\caption{	\label{fig:lyap_taustar_G2osc}
Lyapunov exponent  $\lambda$ in terms of the correlation time  $\tau_*$, for $\varepsilon=1-10^{-4}$. 
Both variables are made dimensioness using  $\lambda_0$
Here  $\overline{\frac{\partial^2 G}{\partial\theta^2}}=0$.
The straight line is a power law with exponent $-1$.
}
\end{figure}

Figure \ref{fig:lyap_taustar_G2osc}  show a plot of $\lambda$.
We find that  $\lambda/\lambda_0 \to 1$
when $\tau_* \to 0$, as in the Markovian case.
For weak noise, and $\tau_*$ large enough \eqref{eq:lyap_G2osc} is well reproduced.

\section{Many degrees of freedom}

\vspace{.5cm}

{\bf Largest exponent (annealed)}

\vspace{.5cm}

We start from equations (\ref{aaa}) and (\ref{bbb}). Putting
$\hat u_{ii}= \sum_{ j}\frac{\partial^2 H }{\partial I_j \partial I_i}	u_{Ij}$ we obtain:
\begin{eqnarray}
\dot u_{\theta i} &=&
\;	\hat u_{Ii}  \label{aaa1}
\\
\dot u_{Ii} &=&  \sum_j { \hat \rho}_{ij}(t) u_{\theta j}
	 \label{bbb1}
\end{eqnarray}
with $\overline{\hat \rho_{ij}(t)\hat \rho_{kl}(t')} = \delta(t-t') \hat \Lambda_{ijkl}$, where:
\begin{equation}
 \hat \Lambda_{ijkl}  = \sum_{ i'm'}\frac{\partial^2 H }{\partial I_i' \partial I_i} \frac{\partial^2 H }{\partial I_m \partial I_m'}
 \Lambda_{i'lm'n}
\end{equation}
Equation (\ref{fp}) becomes:
\begin{equation}
\frac{\partial P}{\partial t}
=
\left[
	\hat u_{Ii} \frac{\partial}{\partial u_{\theta i}}
+
\hat \Lambda_{ijlm}
\frac{\partial^2}{\partial \hat u_{Ii} \partial \hat u_{Ij}}
u_{\theta l} u_{\theta m}
\right]
P
\label{fp1}
\end{equation}
We may generalize the calculation of the largest (annealed) Lyapunov exponent by considering the $3N(N-1)/2$-dimensional vector:
$\{ \langle u_{\theta i} u_{\theta j} \rangle , \langle u_{I i} u_{\theta j} +\langle u_{I j} u_{\theta i} \rangle/2,
 \langle u_{I i} u_{I j} \rangle \}$

\begin{equation}
\frac{d}{dt}
\begin{pmatrix}
\langle u_{\theta i} u_{\theta j} \rangle \\
\langle u_{\theta i} u_{Ij} + u_{\theta j} u_{Ii}\rangle/2 \\
\langle u_{iI} u_{Ij} \rangle \\
\end{pmatrix}
=   
2
\left[
\begin{pmatrix}
0 &1 &0 \\
0 &0 &\frac12 \\
0 &0 &0 \\
\end{pmatrix}
+
\varepsilon
\begin{pmatrix}
0 & 0 & 0 \\
0 & 0 & 0 \\
{\bf \hat \Lambda} & 0 & 0
\end{pmatrix}
\right]
\begin{pmatrix}
\langle u_{\theta i} u_{\theta j} \rangle \\
\langle u_{\theta i} u_{Ij} + u_{\theta j} u_{Ii}\rangle/2 \\
\langle u_{iI} u_{Ij} \rangle \\
\end{pmatrix}
\end{equation}
Just as in the one dimensional case, it is easy to see just by writing the eigenvalue equation, that all eigenvalues satisfy:
\begin{equation}
\frac{1}{2} \varepsilon \hat \Lambda_{ij,kl} V_{kl} = \mu^3 V_{ij}
\end{equation}
The annealed version of the  largest Lyapunov exponent is given by
\begin{equation}
2 \lambda_{max}^{(2)}= \mu_{M}
\end{equation}

\vspace{.5cm}

{\bf Kolmogorov-Sinai entropy}

\vspace{.5cm}

In order to obtain a Riccati form for (\ref{aaa1}) and (\ref{bbb1}), we start by writing them as:
\begin{equation}
\ddot u_{\theta i} 
=  \sum_j { \hat \rho}_{ij}(t) u_{\theta j}
	 \label{abab}
\end{equation}
We now apply this to $N$ independent vectors $u^l_{\theta i}$, which we shall denote as an $N\times N$ matrix $\Theta$:
\begin{equation}
\ddot \Theta_{il} 
=  \sum_j { \hat \rho}_{ij} \Theta_{jl}
	 \label{abab1}
\end{equation}
Defining the {\em matrix} Riccati variable $Z= \dot \Theta \Theta^{-1}$, we get the equation:
\begin{equation}
\dot Z _{ij}- [Z^2]_{ij} = {\hat \rho}_{ij}
\end{equation} 
The Kolmogorov-Sinai entropy is given by the rate of ($N$-dimensional) volume expansion in the $\theta_i$ space:
\begin{equation}
h_{KS}= \langle \frac{d}{dt} \mbox{Tr}\, \ln \Theta \rangle  = \mbox{Tr}\, \langle Z \rangle
\end{equation}
where we have used the identity
$\frac{d}{dt} \mbox{Tr} \ln \Theta= \mbox{Tr}\, \{\dot \Theta \Theta^{-1}\}$

\section{An example:  Foucault's pendulum}

Consider Foucault's pendulum. The one in the Musee des Arts et Metiers
in Paris has a mass of $m=25\;\text{kg}$ with radius $R=0.09\;\text{m}$ hanging at the
end of a $18$~m thread.
The frequency of small oscillations is  $\omega_0=\sqrt{g/l}$ where
$g=9.81\;\text{m}/\text{s}^2$.
The pendulum describes small oscillations of amplitude
$q_0\sim 10^{-2}$ radians.
As we saw above, for a simple pendulum at small oscillations  $\left.\frac{dT}{dH}\right|_{H=0}=\frac{T_o}{8mg}  
$
We neglect  friction, because we assume that some mechanism compensates it. 
On the other hand, we assume the stochastic element of the force fluctuations may be considered
to be Markovian. We are thus led to a situation where  
  $G=-q$, so that  $\overline{(\ddot G)^2} \sim \omega_0^4 q_0^2$.
The diffusion constant in air is related to the viscosity via the Stokes-Einstein relation:
\begin{equation}
D = \frac{k_B T}{6\pi\eta R}
\sim 10^{-16}\; \text{m}^2/\text{s}.
\end{equation}
The intensity of noise is then
$D/(l^2\omega_0)$.
The Lyapunov time is given estimated by  \eqref{eq:tau} 
\begin{equation}
\tau
\sim
\left(
        \frac{D}{l^2\omega_0}
        (\omega_0^4 q_0^2)
        \frac1{\omega_0} \frac1{64}
        \right)^{-1/3}
\sim
\left(\frac{ q_0^2 \omega_0^2 D}{64 \, l^2} \right)^{-1/3}
\end{equation}
We find $\tau \sim 5$ years.
For a pendulum of length in the order of  centimeters, and a mass of
radius in the order of millimeters, the Lyapunov time turns out to be in
the order of days.


\section{Stochastic treatment to model weakly nonintegrable systems beyond KAM regime}

The main motivation of this paper is the perspective  of treating weakly
nonitegrable systems beyond the KAM regime, by substituting the
integrability-breaking interactions by random noise. For example, one might hope to obtain 
an estimate of the Lyapunov exponents of a planet by treating the perturbation due to the other planets as stochastic.

\vspace{.5cm}

{\bf Many-body Lyapunov exponents and {\em passive} approximation}

\vspace{.5cm}

In order to fix ideas, consider a weakly interacting system such as a planetary  system with $N$ planets. In order to test its stability properties
of the orbits of planet $A$
one may proceed in different ways:
\\
 {\em i)} Compute two  trajectories starting with sightly different positions $r_A(t)$ and $r_A(t)+\delta r_A(t)$ of planet $A$, using 
 two copies of the full $6N$-dimensional dynamics $r_i(t)$ and $r_i(t)+\delta r_i(t)$, and then measuring the evolution of the distance  $\delta r_A$ between the two copies of planet $A$.\\
{\em ii)} Compute two nearby trajectories  $r_A(t)$ and $r_A(t)+\delta r_A(t)$ of planet $A$, but treating planet $A$ using the same trajectory of all other planets 
in the two copies (i.e. imposing $\delta r_i(t)=0$ for $i \neq A$) . Planet $A$ is {\em passive} in the sense that the change in its initial conditions and subsequent trajectory does not reflect in a change  in the trajectory of all others.\\
{\em iii)} Even more extreme, one may neglect all interactions except those that the other planets exert on $A$: planet $A$
is then completely passive.

Procedure {\em (i)} gives, for any $A$ and  at long times, the largest Lyapunov exponent of the whole system,  even if separation between trajectories is measured only for planet $A$, although finite-time effects may 
be large and long lasting \cite{tanos}. The reason is easy to understand: the Lyapunov vector has a norm that grows exponentially
with time, and, unless its projection  with any particular direction vanishes  exponentially with time, its time dependence will follow
that of the norm.

Procedures {\em (ii)} and {\em (iii)} give different values for each planet,
and these are approximations that 
for weakly interacting systems might give a  very good estimate of the finite-time sensitivity to initial conditions of a single planet.
One may also conjecture that in that limit the exponents so obtained, treating by turns each planet as passive, might give
a good approximation of the entire set 
of  $6N$  Lyapunov exponents $\lambda_1,...,\lambda_{6N}$, if the exponents are widely different.

For a general system with weak interaction:
\begin{equation}
H= H(I_1,...,I_N) + \epsilon H_{int} (\theta_1,...,\theta_N, I_1,..., I_N)
\end{equation}
with equations of motion
\begin{eqnarray}
\dot \theta_i &=& \omega_i(I_1,...,I_N) + \epsilon \frac{\partial H_{int}}{{\partial I_i}} \nonumber \\
\dot I_i &=& - \epsilon \frac{\partial H_{int}}{{\partial \theta_i}} 
\end{eqnarray}
procedure {\em (iii)} for a single degree of freedom amounts to calculating  the Lyapunov exponent of the following system:
 \begin{eqnarray}
\dot \theta_A&=& \omega_i(I_1,...,I_N) + \epsilon \frac{\partial H_{int}}{{\partial I_A}} \nonumber \\
\dot I_A &=& - \epsilon \frac{\partial H_{int}}{{\partial \theta_A}} 
\end{eqnarray}
where $H_{int}$ is taken as a function of $\theta_A,I_A$ and all other values  $i\neq A$ 
are fixed as $\theta_i=\theta_i(0)+\omega_i t$,
and $I_i(t)=I_i(0)$.
 
\vspace{.5cm}

{\bf Froeschl\'e model}

\vspace{.5cm}

Let  us consider a toy model, which turns out to be quite instructive. We study the $N$ degree of freedom
version of the Hamiltonian introduced in \cite{froeschle2000}:
\begin{equation}	\label{eq:froeschle:H}
H_F = \sum_{i=1}^N \frac{I_i^2}{2} + I_0
    + \frac{\epsilon (N+2)}{\displaystyle 1 + \frac1{N+2} \sum_{i=0}^N \cos
\theta_i}
\end{equation}
We have scaled coefficients so that both terms are extensive and
$\epsilon$ intensive.
When  $\epsilon=0$, the system is integrable, and the  $\theta_i$
turn with angular speed  $I_i$, except for  $\theta_0$, which has unit angular frequency.
When $\epsilon>0$, $H$ is no longer integrable.
The equations of motion read, with $i\ge 1$,
\begin{align}
\dot \theta_0 &= 1, \nonumber\\
\dot \theta_i &=  I_i \nonumber\\
\dot I_i &= -\epsilon\sin{\theta_i} - \epsilon\sin{\theta_i} \xi(t)
,
\label{f1}
\end{align}
with
\begin{equation}
\xi(t) = \left(1 + \frac1{N+2} \sum_i \cos\theta_i \right)^{-2} - 1
.
\end{equation}

If $\epsilon>0$ is small enough, the KAM theorem applies 
and some invariant tori survive. The values of $\epsilon$ for this to be the case are expected to be
exponentially small in $N$, and become extremely small already for $N=6$ \cite{tailleur2007}.
 The first regions of phase space where tori break, are the places where
$\sum_i n_i \omega_i = 0$,
where the  $n_i$ are integers.
This scenario was observed in Ref. \cite{froeschle2000} for  $N=2$.

We shall consider a torus given by $I_i (= \omega_i)$  chosen from a Gaussian  distribution
 with zero mean and variance
 $\beta^{-1}$.
{\em For $I_i$ incommensurate},  
the quantity $\xi(t)$ is a sum of projections of incommensurate angles, and one expects it to behave as 
a pseudo-random number generator,  at least for $N$ large enough. 
The question as to if and when such signals may be taken as random, and the more refined one of the recurrences in their autocorrelations, has received enormous attention both in mathematics and physics. (The reader will find a discussion
and references in Zwanzig's book \cite{zwanzig1}).

\vspace{.5cm}

{\bf Statistical properties of $\xi(t)$}

\vspace{.5cm}

If we assume that the angles are random enough that $\sum_{i=0}^N \cos{\theta_i}=O(\sqrt(n))$ , we may develop for large $N$ :
\begin{equation}
\xi(t) =
	-\frac2{N+2} \sum_{i=0}^N\cos{\theta_i}
	+\frac3{(N+2)^2} \left(\sum_{i=0}^N \cos{\theta_i}\right)^2
	+ {\cal O}(N^{-3})
.
\end{equation}
The assumption that the $\theta_i$ are decorrelated angles requires at
the very least that we are not on a resonance.  This amounts to treating
$\xi(t)$ as deriving from $\theta_i$ that are independent, random, and
uniformly distributed in $[0, 2\pi]$.  To lowest order, we have:
\begin{align}		\label{eq:froeschle:avgxi}
\langle \xi \rangle
&=
\frac32 \frac{N+1}{(N+2)^2}
\simeq \frac{3}{2N}, \\
\langle \xi^2 \rangle
&=
2 \frac{N+1}{(N+2)^2}
\simeq \frac{2}{N}
\end{align}
and 
\begin{equation}	\label{sigma2estimate}
\sigma^2
=
\langle\xi^2\rangle-\langle\xi\rangle^2
\simeq \frac2N
.
\end{equation}
Let us now calculate the autocorrelation of  $\xi(t)$.
We consider constant $\omega_i$ ($=I_i$), as we shall be interested in the dynamics before the system
leaves the vicinity of a torus. We hence put $\theta_i(t) = \theta_i(0) + \omega_i t$.
For large values of  $N$,
\begin{equation}
C(t) = 
\frac{1}{\sigma^2}
\langle \xi(0) \xi(t) \rangle
=
\frac{4}{\sigma^2(N+2)^2} \sum_{i,j}
	\langle \cos\bigl[\theta_i(0)\bigr] \cos\bigl[\theta_i(0)+\omega_i t\bigr] \rangle
=
\frac{1}{N+1} \sum_{i} \cos{\omega_i t}	
.
\end{equation}
We have averaged over initial conditions: $\theta_i(0)$  drawn from a uniform distribution in $[0,2\pi]$.
We wish to estimate the time  $\tau_*$ 
of decay of the correlation.
We have $C(0) = 1$ by construction, and as 
 $t\gg\tau_*$, if the  $\omega_i$ are not commensurable,
$C(t)$ is a fluctuating quantity with variance $1/N$.

\begin{figure}
\begin{center}
\includegraphics[width=5cm]{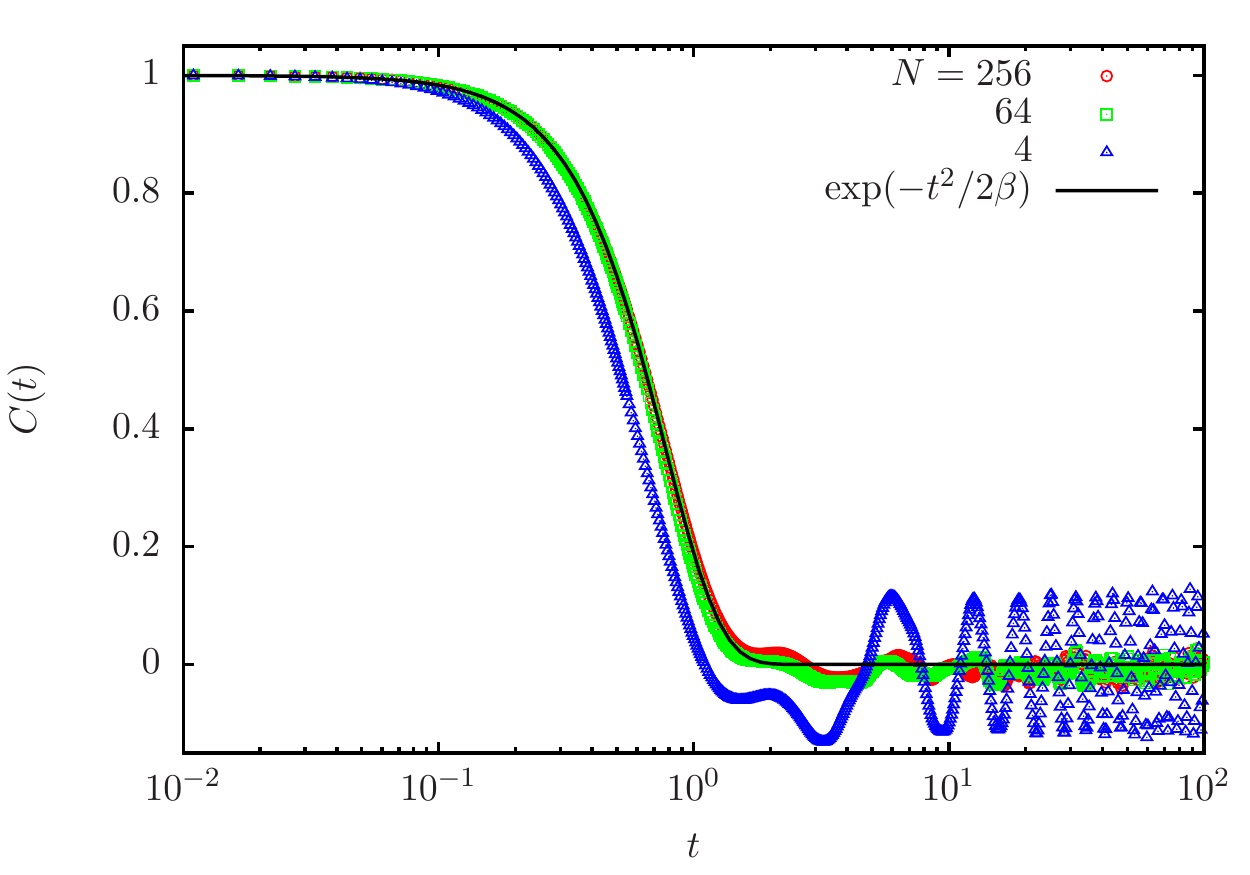}\\
\end{center}
\caption{	\label{stochtest0}
Perurbation autocorrelation}
\end{figure}

The value of  $\tau_*$ depends on the distribution of the
$\omega_i$.
For large $N$,  we have
\begin{equation}
C(t) =
\frac1{\sqrt{2\pi\beta^{-1}}}
	\int \cos{\omega t} \; e^{-\beta{\omega^2}/{2}} d\omega 
=
\exp\left(-\frac12 \beta^{-1} t^2\right)
,
\end{equation}
so that
\begin{equation}	\label{tauestimate}
\tau_* = \beta^{1/2}
\end{equation}
may be interpreted as the autocorrelation time. It is of the order of
the average period of oscillation, and is independent of $N$. The autocorrelation
is shown in Fig \ref{stochtest0}

In principle, one could make some effort to calculate better estimates
of $\sigma^2$ and $\tau_*$.  However, the power law with exponent
$\frac13$ in the scaling law~\eqref{scalingtest} tells us that the
Lyapunov exponent only depends little on  these parameters, so that
very precise estimates are not needed.

\vspace{.5cm}

{\bf Testing validity of stochastic treatment.}

\vspace{.5cm}

Let us first consider an extreme form of `passive' approximation: we shall see how $\xi(t)$ acts on a single degree
of freedom that has no feedback on the rest of the variables:
\begin{equation}	\label{eq:bruitmetre}
(I)\qquad
\begin{aligned}
\dot u_\theta &= u_I \\
\dot u_I &= \varepsilon^{1/2} \xi(t) \, u_\theta
\end{aligned}
\end{equation}
If $\xi(t)$ is a real noise with zero average, variance
$\langle \xi^2(t)\rangle=\sigma^2$ and correlation time $\tau_*$, we
know from the results of section~\ref{sec:1dof} that the Lyapunov
exponent of this dynamics should scale as
\begin{equation}	\label{scalingtest}
\lambda \propto \left(\varepsilon \sigma^2 \tau_* \right)^{1/3}
.
\end{equation}
Note that $\varepsilon$ is a parameter (not to be confused
with the perturbation parameter $\epsilon$) that can be freely varied,
providing some flexibility in testing the validity of our stochastic
treatment.

\begin{figure}
\begin{center}
\includegraphics{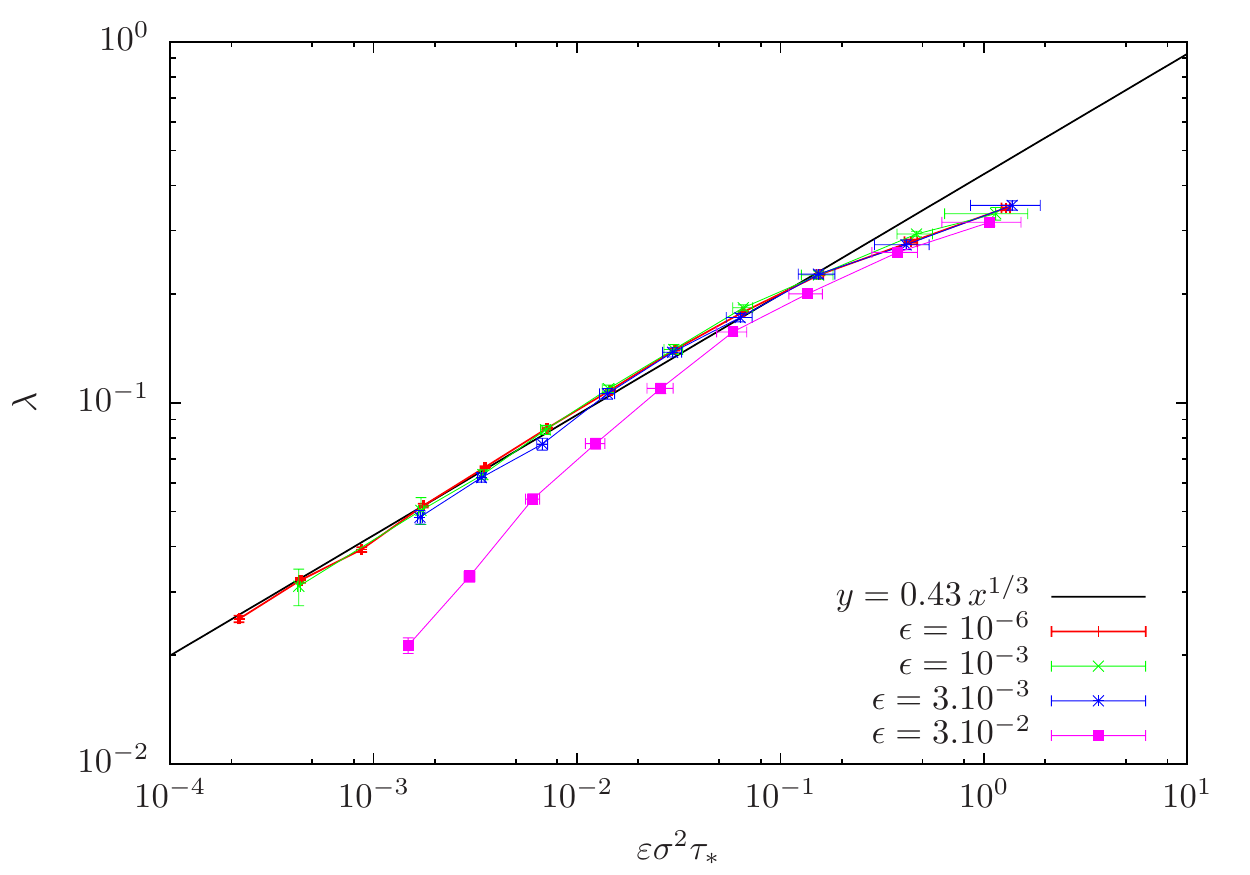}\\
\end{center}
\caption{	\label{stochtest}
Measured Lyapunov exponent of the dynamics~\eqref{eq:bruitmetre} for
$\varepsilon=1$ and
$\beta=0.8$.
The `noise' $\xi(t)$ was generated for different values of $\epsilon$
and $N$ in the Froeschle dynamics~\eqref{eq:froeschle:H}.
The variance $\sigma^2$ was measured numerically and $\tau_*$ was
estimated using~\eqref{tauestimate}.
Each point is an average over realisations of different initial conditions.
}
\end{figure}

We now substitute the gaussian noise by the one generated in a true Froeschl\'e model
setting (artificially) the parameter $\varepsilon$ to one. 
We used different realisations of $\xi(t)$ for $\beta=0.8$, $N$ between
$4$ and $8192$ and $\epsilon$ of order $10^{-3}$ (see Fig~\ref{stochtest}).
We measured the Lyapunov exponent by estimating the exponential rate of
growing of $(u_\theta,u_I)$.

Figure \ref{stochtest} shows the Lyapunov exponent obtained in this way, compared to the analytical expression,
with the autocorrelation time $\tau_*$  estimated from~\eqref{tauestimate}.
The agreement is very good for weak values of noise, over several decades of noise intensity.

\vspace{.5cm}

{\bf The problem of an unusually unstable degree of freedom:  the   Modified Froeschl\'e model}

\vspace{.5cm}

Next step is to go back to the true Froeschl\'e model and compute its true Lyapunov exponent, and compare it with the largest one obtained by procedure {\em (iii)} above, that is, by mimicking the effect on a given degree of freedom 
by mimicking the perturbation produced by all other degrees of freedom  with a random correlated noise. 
 Because the effective perturbation  has amplitude $O(N^{-\frac12})$, one would expect the Lyapunov exponent to be proportional
 to $N^{-\frac13}$.  We have tried and {\em we have checked numerically that this is not so}, the largest Lyapunov exponent 
 is very weakly dependent of $N$, if at all.
 
 The explanation of this surprising fact is instructive. In our procedure we are choosing $N$ values of $I_i$ in an interval $\beta$ of order one.  Each degree of freedom performs a motion following equation \ref{f1}, corresponding to a pendulum of 
 amplitude with `energy' $I_i^2/2$ and `gravity field'  $\sim \epsilon$ (because of the term $-\epsilon \sin \theta_i$ in the last equation).  Some  of these degrees of freedom are close to the separatrix,  
 the distance being 
 \begin{equation}
 \delta_i = \left| \frac{I_i^2}{2}-\epsilon\right|
 \end{equation}
 One may estimate from the Gaussian distribution of the $I-i$ that the smallest $\delta_i$ scale as 
 $\delta_{min} \sim \frac{1}{N \epsilon}$, while the noise scales as $\varepsilon_{eft} \sim N^{-\frac 12}$. 
Recall now the discussion of section \ref{sepa}: the Lyapunov exponent of a degree of freedom scales
as $\varepsilon_{eff} |\ln \delta_i|^2 \delta_{-\frac 23}$, which actually {\em increases} with $N$.
 Because the global Lyapunov exponent, whichever projection we measure, will be dominated by the largest,
 we conclude that the exponent is much larger than $\sim N^{\frac 13}$.
 In a word, the "crowding" of many degrees of freedom  as $N \rightarrow \infty$ has produced interactions that become large, and in fact grow with  $N$.

 On way to minimize this problem is consider a model with an extra term:
\begin{equation}	\label{eq:froeschle:H}
H_{mF} = \sum_{i=1}^N \frac{I_i^2}{2} + I_0
    + \frac{\epsilon (N+2)}{\displaystyle 1 + \frac1{N+2} \sum_{i=0}^N \cos \theta_i}+\sum_{i=0}^N \cos
\theta_i
\end{equation}
so that the  equations of motion read, with $i\ge 1$,
\begin{align}
\dot \theta_0 &= 1, \\
\dot \theta_i &=  I_i \\
\dot I_i &=  - \epsilon\sin{\theta_i} \xi(t)
,
\label{f2}
\end{align}

\begin{figure}
\begin{center}
\includegraphics{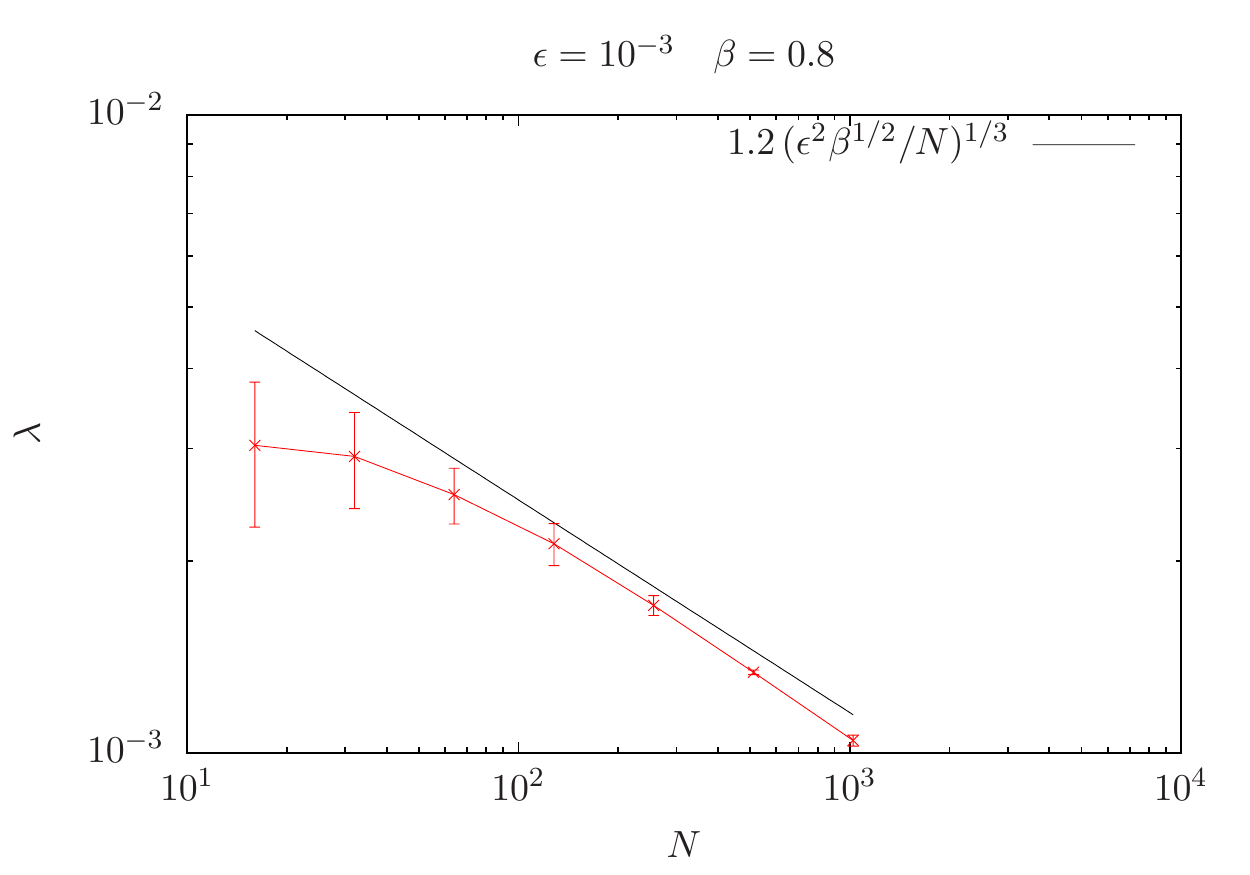}\\
\end{center}
\caption{	\label{stochtest1}
Measured Lyapunov exponent of {\em one degree of freedom} (see text) of the modified Froeschle model
$\varepsilon=1$ and
$\beta=0.8$.
}
\end{figure}

We have measured the lyapunov exponent of one {\em passive} degree of freedom of this model,
and obtained a good agreement even for relatively low values of $N$ (see Fig \ref{stochtest1})

\section{Conclusions}

We have derived expressions for the Lyapunov exponents of an integrable system perturbed by additive stochastic
noise.  The motivation is to use this knowledge to estimate the effect of deterministic perturbations on an almost integrable
system which is, however, far from the KAM and Nekhoroshev regimes -- as will be the case as soon as one considers 
systems with many degrees of freedom and reasonable strong perturbations. 
The field of application of such an approach could be widespread: we have mentioned already planets and stellar clusters,
but even a sound wave traveling in a liquid is an example of a near integrable system interacting with many (microscopic) degrees of freedom.
 
 Already at a phenomenological level the approach allows us too understand some global features of weakly chaotic (but far from KAM!)
 systems. As an example, consider the Fermi-Pasta-Ulam ($\alpha+\beta$) chain, which as argued in Ref. \cite{benettin} may be considered
 as an integrable Toda lattice plus an integrability breaking perturbation. The long thermalization time is attributed to the near-integrability, 
 the motion is fast within a (Toda) torus, and slowly drifts between tori. However, a surprise appears when one computes the Lyapunov instability: it turns out that the Lyapunov time is much shorter than the thermalization time, and indeed scales differently on approaching the  
 Today limit \cite{benettin}. The result in this paper allows to guess the resolution of this paradox: most of the Lyapunov instability
 is expected to happen  tangentially to the tori, and the effect of diffusion away from them is expected to be of higher order.
 Indeed, we recognize the same situation of planets in the solar system, which are enormously stable as compared to their Lyapunov times --
 for exactly the same reason.
 
 We suppose  that an approximation of integrability-breaking terms as random noise must be implicitly present in one way or another in the literature (see e.g. \cite{astrid}),
 but a systematic and general study seems to be missing. We hope this paper may offer a step in that direction.

\vspace{.5cm}

{\bf \large Acknowledgments}

\vspace{.5cm}

We wish to thank  G. Benettin, M. Chertkov,  A. Politi, S. Ruffo, S. Tremaine and A. deWijn for clarifying remarks and suggestions.



\begin{thebibliography}{99}

\bibitem{celia} The exponent $\frac{1}{3}$ is also familiar in the theory of products of random matrices, see:
C Anteneodo and R. O. Vallejos. Phys Rev {\bf  E} 65.1 (2001): 016210;
 Physical Review E 85.2 (2012): 021124.

\bibitem{aris1956}
R.~{Aris},
 {\em On the Dispersion of a Solute in a Fluid Flowing through a Tube},
 {Royal Society of London Proceedings Series A} 235:67--77, 1956.

\bibitem{cepas1998}
O.~C{\'e}pas and J.~Kurchan,
 {\em Canonically invariant formulation of {L}angevin and {F}okker-{P}lanck equations.}
 {EPJ B} 2(2):221--223, 1998.

\bibitem{Derrida}
B.~{Derrida} and E.~{Gardner},
 {\em Lyapounov exponent of the one dimensional anderson model: weak disorder expansions},
 {Journal de Physique} 45:1283--1295, 1984.

\bibitem{FGP}
S.~{Fishman}, D.~R. {Grempel}, and R.~E. {Prange},
 {\em Chaos, Quantum Recurrences, and Anderson Localization},
 {Phys. Rev. Lett.} 49:509--512, 1982.

\bibitem{froeschle2000}
C.~Froeschlé, M.~Guzzo, and E.~Lega,
 {\em Graphical evolution of the arnold web: from order to chaos},
 {Science} 289(5487):2108--2110, 2000.

\bibitem{gardiner}
C.W. Gardiner,
 {\em Handbook of Stochastic Methods},
 Springer, 2nd edition, 1985.

\bibitem{guyon}
\'E. Guyon, J.-P. Hulin, and L.~Petit,
 {\em Hydrodynamique physique},
 EDP Sciences, 3rd edition, 2012.

\bibitem{Halperin}
B.~I. {Halperin},
 {\em Green's Functions for a Particle in a One-Dimensional Random Potential},
 {Physical Review} 139:104--117, 1965.

\bibitem{Landau}
L.~D. {Landau} and E.~M. {Lifshitz},
 {\em Quantum mechanics}.

\bibitem{laskar1989}
J.~Laskar,
 {\em A numerical experiment on the chaotic behaviour of the solar system},
 {Nature} 338:237--238, 1989.

\bibitem{Livi}
R.~{Livi}, M.~{Pettini}, S.~{Ruffo}, M.~{Sparpaglione}, and A.~{Vulpiani},
 {\em Equipartition threshold in nonlinear large Hamiltonian systems: The Fermi-Pasta-Ulam model},
 {Phys. Rev. A} 31:1039--1045, 1985.

\bibitem{mallick2002}
K.~{Mallick} and P.~{Marcq},
 {\em Anomalous diffusion in nonlinear oscillators with multiplicative noise},
 {Phys. Rev. E} 66(4):041113, 2002.

\bibitem{Morbi}  See, for example, the discussion in:  
A.~{Morbidelli} and C.~{Froeschl{\'e}},
 {\em On the Relationship Between Lyapunov Times and Macroscopic Instability Times},
 {\em Celestial Mechanics and Dynamical Astronomy} 63:227--239, 1996.


\bibitem{misha} M Chertkov, I. Kolokolov, V. Lebedev and K. Turistin,
J. Fluid Mechanics 531 (2005),  251-260


\bibitem{risken}
H.~Risken,
 {\em The Fokker-Planck Equation: Methods of Solution and Applications},
 Springer-Verlag, 2nd edition, 1989.

\bibitem{schomerus2002}
H.~Schomerus and M.~Titov,
 {\em Statistics of finite-time {L}yapunov exponents in a random time-dependent potential},
 {Phys. Rev. E} 66:066207, 2002.

\bibitem{sussmanwisdom}
G.~J. Sussman and J.~Wisdom,
 {\em Chaotic evolution of the solar system},  
 {Science} 257(5066):56--62, 1992;
%
G.~J. {Sussman} and J.~{Wisdom},
 {\em Numerical evidence that the motion of Pluto is chaotic},
 {Science} 241:433--437, 1988.

\bibitem{tailleur2007}
J.~{Tailleur} and J.~{Kurchan},
 {\em Probing rare physical trajectories with {L}yapunov weighted dynamics},
 {Nature Physics} 3:203--207, 2007.

\bibitem{Tannor}
David~J. {Tannor},
 {\em Introduction to quantum mechanics},
 University Science Books, 2007.

\bibitem{taylor1953}
G.~{Taylor},
 {\em Dispersion of Soluble Matter in Solvent Flowing Slowly through a Tube},
 {Royal Society of London Proceedings Series A} 219:186--203,
  1953.

\bibitem{astrid} A. S. de Wijn, B. Hess, B. V. Fine
Phys. Rev. Lett. 109, 034101 (2012) ;  arXiv:1209.1468.


\bibitem{tanos} Campa, A., A. Giansanti, and A. Tenenbaum,  J Phys {\bf A}: Math and Gen (1999): 1915.

\bibitem{tessieri2000}
L.~Tessieri and F.~M. Izrailev,
 {\em Anderson localization as a parametric instability of the linear kicked oscillator},
 {Phys. Rev. E} 62(3):3090, 2000.

\bibitem{pendu_p}
\url{http://en.wikipedia.org/wiki/Pendulum}

\bibitem{wisdom1987}
J.~{Wisdom},
 {\em Urey Prize Lecture: Chaotic dynamics in the solar system},
 {Icarus} 72:241--275, 1987.

\bibitem{Zwanzig}
R.~{Zwanzig},
 {\em Nonlinear generalized Langevin equations},
 {J. Stat. Phys.} 9:215--220, 1973.


\bibitem{japoneses} 	
Hondou, Tsuyoshi, and Yasuji Sawada Physical review letters 75.18 (1995): 3269-3272. 


\bibitem{zwanzig1} 	
R. Zwanzig, {\em Nonequilibrium statistical mechanics} . Oxford University Press, USA, 2001.

\bibitem{benettin}Benettin, G., and A. Ponno, Journal of Statistical Physics 144.4 (2011): 793-812.

\end{thebibliography}
\end{document}